\documentclass[11pt]{article}
\pdfoutput=1
\usepackage[centertags]{amsmath}
\usepackage[square, comma, sort&compress,numbers]{natbib}
\usepackage{array,multirow}
\numberwithin{equation}{section}
\usepackage{amssymb,mathtools}
\usepackage{amsfonts,amsthm,stmaryrd}
\usepackage{graphicx}
\usepackage{color,bm}
\usepackage[normalem]{ulem}

\usepackage{epsfig}
\usepackage{bbold}
\usepackage{subfigure}

\usepackage{wrapfig}
\usepackage{float}
\usepackage{soul}
\usepackage{color}
\usepackage{ulem}

\usepackage{comment}

\usepackage[dvipsnames]{xcolor}

\usepackage{tikz,pgf}
\usetikzlibrary{shapes}
\usetikzlibrary{calc}
\usetikzlibrary{decorations.pathmorphing}
\usetikzlibrary{decorations.pathreplacing,shapes.misc}
\usetikzlibrary{positioning}
\usetikzlibrary{arrows}
\usetikzlibrary{decorations.markings}
\usetikzlibrary{shadings}

\usetikzlibrary{intersections}

\def\sideremark#1{\ifvmode\leavevmode\fi\vadjust{\vbox to0pt{\vss
 \hbox to 0pt{\hskip\hsize\hskip1em
 \vbox{\hsize3cm\tiny\raggedright\pretolerance10000
  \noindent #1\hfill}\hss}\vbox to8pt{\vfil}\vss}}}

\newcommand{\be}{\begin{equation}}
\newcommand{\ee}{\end{equation}}
\newcommand{\ba}{\begin{eqnarray}}
\newcommand{\ea}{\end{eqnarray}}

\def\tr{{\rm Tr}}

\advance\textheight\topskip
\renewcommand{\tilde}{\widetilde}

\newcommand{\binner}[2]{%
  {\langle}\kern-4.15pt{\langle}#1{,}\,#2{\rangle}\kern-4.15pt{\rangle}}

\newcommand{\ffrac}[2]{\raisebox{.5pt}%
  {\footnotesize$\displaystyle\frac{#1}{#2}$}\kern1pt}

\numberwithin{equation}{section} \makeatletter
\@addtoreset{equation}{section}

\def\be{\begin{equation}}
\def\ee{\end{equation}}
\def\ba{\begin{array}}
\def\ea{\end{array}}

\def\tr{{\rm Tr}}

\newdimen\tableauside\tableauside=1.0ex
\newdimen\tableaurule\tableaurule=0.4pt
\newdimen\tableaustep
\def\phantomhrule#1{\hbox{\vbox to0pt{\hrule height\tableaurule
width#1\vss}}}
\def\phantomvrule#1{\vbox{\hbox to0pt{\vrule width\tableaurule
height#1\hss}}}
\def\sqr{\vbox{%
  \phantomhrule\tableaustep

\hbox{\phantomvrule\tableaustep\kern\tableaustep\phantomvrule\tableaustep}%
  \hbox{\vbox{\phantomhrule\tableauside}\kern-\tableaurule}}}
\def\squares#1{\hbox{\count0=#1\noindent\loop\sqr
  \advance\count0 by-1 \ifnum\count0>0\repeat}}
\def\tableau#1{\vcenter{\offinterlineskip
  \tableaustep=\tableauside\advance\tableaustep by-\tableaurule
  \kern\normallineskip\hbox
    {\kern\normallineskip\vbox
      {\gettableau#1 0 }%
     \kern\normallineskip\kern\tableaurule}%
  \kern\normallineskip\kern\tableaurule}}
\def\gettableau#1 {\ifnum#1=0\let\next=\null\else
  \squares{#1}\let\next=\gettableau\fi\next}

\def\cA{\mathcal{A}}

\def\cD{\mathcal{D}}

\def\cL{\mathcal{L}}
\def\cM{\mathcal{M}}

\def\cO{\mathcal{O}}
\def\cP{\mathcal{P}}

\numberwithin{equation}{section} \makeatletter
\@addtoreset{equation}{section}

\newcommand{\hol}{\text{Hol} }

\def\cft1{\text{CFT}_{1}}

\def\be{\begin{equation}}
\def\ee{\end{equation}}
\def\ba{\begin{array}}
\def\ea{\end{array}}

\def\ba{\begin{array}}
\def\ea{\end{array}}

\def\ie{i.e.}

\numberwithin{equation}{section} 
\numberwithin{table}{section} 

\usepackage{sectsty} 
\allsectionsfont{\centering } 

\def\XXint#1#2#3{{\setbox0=\hbox{$#1{#2#3}{\int}$}
     \vcenter{\hbox{$#2#3$}}\kern-.5\wd0}}

\def \tr{\mbox{tr\,}}

\usepackage{jheppub}

\makeatletter
\def\@fpheader{\vspace{-.1cm}}
\makeatother

\title{\centering{ Gravitational Edge Mode in Asymptotically AdS$_2$\\: JT Gravity Revisited}}

 \author[a]{Euihun Joung}
 \author[b]{Prithvi Narayan}
 \author[c,d,e]{Junggi Yoon}

\affiliation[a]{Department of Physics, Kyung Hee University Seoul 02447, Korea
}
\affiliation[b]{Department of Physics, Indian Institute of Technology, Palakkad 678557, India}
\affiliation[c]{Asia Pacific Center for Theoretical Physics,\\77 Cheongam-ro, Nam-gu, Pohang-si, Gyeongsangbuk-do, 37673, Korea}
\affiliation[d]{Department of Physics, POSTECH\\ 77 Cheongam-ro, Nam-gu, Pohang-si, Gyeongsangbuk-do, 37673, Korea}
\affiliation[e]{School of Physics, Korea Institute for Advanced Study\\
85 Hoegiro Dongdaemun-gu, Seoul 02455, Korea}

\emailAdd{euihun.joung@khu.ac.kr}
\emailAdd{prithvi.narayan@gmail.com}
\emailAdd{junggi.yoon@apctp.org}

\abstract{We study the gravitational edge mode of the Jackiw-Teitelboim (JT) gravity and its $sl(2,\mathbb{R})$ BF theory description with the asymptotic AdS$_2$ boundary condition. We revisit the derivation of the Schwarzian theory from the wiggling boundary as an action for the gravitational edge mode. We present an alternative description for the gravitational edge mode from the metric fluctuation with the fixed boundary, which is often referred as ``would-be gauge mode''. We clarify the relation between the wiggling boundary and the would-be gauge mode. We demonstrate a natural top-down derivation of $PSL(2,\mathbb{R})$ gauging and the path integral measure of the Schwarzian theory. In the $sl(2,\mathbb{R})$ BF theory, we incorporate the gravitational edge mode and derive the Schwarzian theory with $PSL(2,\mathbb{R})$ gauging. We also discuss the path integral measure from the Haar measure in the Iwasawa decomposition of $PSL(2,\mathbb{R})$.}

\begin{document}

\maketitle

\flushbottom

\section{Introduction}
\label{sec: introduction}

In recent years, the Jackiw-Teitelboim~(JT) gravity with negative cosmological constant has been spotlighted in the study of the SYK model~\cite{Sachdev:1992fk,Polchinski:2016xgd,Jevicki:2016bwu,Maldacena:2016hyu,Jevicki:2016ito,Fu:2016vas,Davison:2016ngz,Murugan:2017eto,Narayan:2017hvh} and its holographic dual~\cite{Almheiri:2014cka,Jensen:2016pah,Maldacena:2016upp,Cvetic:2016eiv,Das:2017pif,Mandal:2017thl,Goel:2020yxl,Bak:2023zkk}, the black hole information paradox~\cite{Almheiri:2019psf,Almheiri:2019hni,Almheiri:2019qdq,Penington:2019kki,Almheiri:2020cfm,Bak:2020enw,Bak:2021qbo}, and the relation to random matrix theory~\cite{Cotler:2016fpe,Saad:2019lba,Stanford:2019vob}. The JT gravity is reduced to the Schwarzian theory on the boundary of AdS$_2$, and the remarkable properties of the Schwarzian theory such as $PSL(2,\mathbb{R})$ gauging~\cite{Maldacena:2016upp,Alkalaev:2022qfc}, the path integral measure~\cite{Bagrets:2016cdf,Stanford:2017thb,Cotler:2018zff,Saad:2019lba} and the one-loop exactness~\cite{Stanford:2017thb} enables analytic control in the full quantum gravity.

The JT gravity is so simple that, at first glance, no physical degree of freedom is left except for the topological ones. Hence, the ``nearly-AdS$_2$'', which allows a wiggling boundary, was introduced to account for the fluctuation on the boundary of AdS$_2$~\cite{Maldacena:2016upp}. This seems to make it distinct from the previous approach to  higher dimensional gravity theories. However, this ``new'' approach with a wiggling boundary brought about questions in the derivation of the Schwarzian theory. For example, the path integral of the JT gravity should include the integration over the wiggling boundary because the resulting Schwarzian theory does. In essence, the wiggling of the boundary is treated as a degree of freedom of the theory, and consequently the variation of the JT action must take the variation of the wiggling boundary into account. And the well-posed path integral with a wiggling boundary should have provided a top-down derivation of the $PSL(2,\mathbb{R})$ gauging and the path integral measure of the Schwarzian theory. Furthermore, it is not clear whether the ``nearly-AdS$_2$'' is fundamentally different from the usual approach to the asymptotically AdS~\cite{Brown:1986nw}.

In fact, the wiggling boundary has been discussed to deal with the broken radial diffeomorphism on the boundary of AdS~\cite{Navarro-Salas:1998fgp,Skenderis:1999nb,Bautier:1999ic,Bautier:2000mz,Rooman:2000ei,Carlip:2005tz,Carlip:2022fwh,Choi:2023syx}. One way to resolve this issue on the breaking of the diffeomorphism is to disallow such broken gauge transformations on the boundary. Since a part of the gauge symmetry is prohibited, we regain additional physical degrees of freedom, so-called \textit{would-be gauge mode}, which would have been gauged away by a radial diffeomorphism if it is not broken. Alternatively, one may introduce a Stueckelberg field to recover the full diffeomorphism invariance on the boundary. And the Stueckelberg field can be ``eaten'' by the field-dependent coordinate transformation, which leads to the wiggling boundary. Therefore, the would-be gauge mode is resurrected as a wiggling boundary.

Exactly the same phenomenon has been intensively studied in the context of the fractional quantum Hall effect~(FQHE) context~\cite{Wen:1989mw,Wen:1990qp,Wen:1990se,Stone:1990by}. For example\footnote{See also~\cite{Balachandran:1994up,Balachandran:1995qa,Carlip:1995cd,Arcioni:2002vv,Blommaert:2018rsf,Blommaert:2018oue,Takayanagi:2019tvn,Donnelly:2020teo,Jiang:2020cqo,David:2022jfd,Mertens:2022ujr,Wong:2022eiu} for other examples of studies on the edge mode.}, in the $U(1)$ Chern-Simons theory on a manifold with boundary, the $U(1)$ gauge symmetry is broken. And, the would-be gauge mode from this broken $U(1)$ gauge symmetry is known as the edge mode described by the chiral boson on the boundary. The JT gravity with the asymptotic AdS$_2$ boundary condition can also be investigated by the $sl(2,\mathbb{R})$ BF theory~\cite{Fukuyama:1985gg,Chamseddine:1989wn,Chamseddine:1989yz,Grumiller:2017qao,Mertens:2018fds,Blommaert:2018iqz,Valach:2019jzv,Ferrari:2020yon,Alkalaev:2022qfc}. Therefore, the Schwarzian theory can be derived as a gravitational edge mode of the $sl(2,\mathbb{R})$ BF theory. In this paper, we will revisit the derivation of the Schwarzian theory from the $sl(2,\mathbb{R})$ BF theory. In particular, we clarify the boundary condition and demonstrate a natural top-down derivation of $PSL(2,\mathbb{R})$ gauging and the path integral measure.

This paper aims at understanding the Schwarzian theory as a gravitational edge mode of the asymptotically AdS$_2$ with $PSL(2,\mathbb{R})$ gauging and the path integral measure from the three different points of view: the wiggling boundary, the would-be gauge mode and the $sl(2,\mathbb{R})$ BF theory. This paper is organized as follows. In Section~\ref{sec: u1 cs theory}, we review the edge mode in the $U(1)$ Chern-Simons theory. In Section~\ref{sec: wiggling boundary}, we derive the Schwarzian theory with a wiggling boundary. In Section~\ref{sec: stueckelberg field}, we also derive the Schwarzian theory from the would-be gauge mode with a fixed boundary, and we present the relation to the wiggling boundary. In Section~\ref{sec: sl2 bf gravity}, we discuss the derivation of the Schwarzian theory from the $sl(2,\mathbb{R})$ BF theory. In Section~\ref{sec: conclusion}, we make concluding remarks.

\section{Review: Edge Mode in $U(1)$ Chern-Simons Theory}
\label{sec: u1 cs theory}

We begin with a brief review of the edge mode in the $U(1)$ Chern-Simons theory on the manifold $\cM\,=\,\mathbb{R}^2\times \{\, x^2\, |\,x^2\geqq 0\, \}$.
\begin{align}
    S_{CS}[A]\,=\, \int_{\cM} \; A\wedge dA\ . 
\end{align}
The variation of the action $S_{CS}$ is 
\begin{align}
    \delta S_{CS}\,=\, 2\int_{\cM } (\delta A \wedge dA) + \int_{\partial \cM} d^2x\;  \big( A_1 \delta A_0- A_0 \delta A_1 \big) \ .
\end{align}
The boundary condition for the well-defined variational principle is usually chosen to be
\begin{align}
    A_0- vA_1\,=\, 0\hspace{10mm}\mbox{on } \partial\cM \,=\, \{ x^1| x^2 \,=\,0\}\ .
\end{align}
In this review, we consider more generic boundary condition
\begin{align}
    A_0-vA_1\;:\;\;\mbox{fixed}\ .
\end{align}
For this boundary condition, we add a boundary term such that the variational principle is well defined as we will see soon
\begin{align}
    S_{bdy}\,=\, \int_{\partial \cM}d^2x\;   A_1(A_0-v A_1)\ .\label{eq: bdy term for cs}
\end{align}
Then, the variation of the total action $S_{tot}\equiv S_{CS}+S_{bdy}$ becomes
\begin{align}
    \delta S_{tot}\,=\, 2\int_{\partial \cM} d^2x \; \big[ A_1\delta (A_0-v A_1) \big]\ .
\end{align}
And we choose the boundary condition\footnote{There is alternative boundary condition $A_1\big|_{x^2=0}=0$.} of $A$ on $\partial\cM$ to be
\begin{align}
    A_0-v A_1\,\big|_{x^2=0}\,=\, J \ .
\end{align}
where $J=J(x^0,x^1)$ is a fixed function.

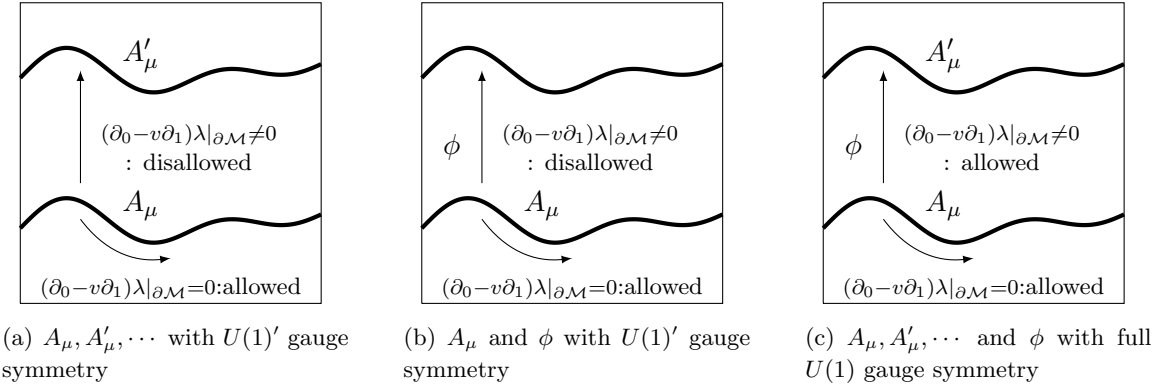
\begin{figure}[t!]
\centering
\subfigure[$ A_\mu,A'_\mu,\cdots$ with $U(1)'$ gauge symmetry]{
\centering
\begin{minipage}[c]{0.28\linewidth}
\centering
\begin{tikzpicture}

\def\length{4}
\def\separation{5}

\draw[thin] (0,0) -- (\length,0) -- (\length,\length) -- (0,\length)--(0,0);

\draw[domain=0:\length,samples=200,smooth, variable=\x,ultra thick] plot ({\x},{1.0+0.1*sin(deg(0.3*\x))+0.2*sin(deg(2*\x))+0.2*sin(deg(3*\x))});

\draw[domain=0:\length,samples=200,smooth, variable=\x,ultra thick] plot ({\x},{3.0+0.1*sin(deg(0.3*\x))+0.2*sin(deg(2*\x))+0.2*sin(deg(3*\x))});

\node[] at ({0.4*\length},1.3) {$A_\mu$};

\node[] at ({0.5*\length},0.2) {$\scriptstyle (\partial_0 -v \partial_1)\lambda|_{\partial\cM}=0${\footnotesize:allowed}};

\draw[-latex] ({0.2*\length},{0.28*\length}) to[bend right] ({0.5*\length},{0.15*\length});

\node[] at ({0.4*\length},3.3) {$A'_\mu$};

\draw[-latex] ({0.2*\length},1.6) -- ({0.2*\length},3.1); 

\node[] at ({0.56*\length},{0.52*\length}) {\shortstack{$\scriptstyle (\partial_0 -v \partial_1)\lambda|_{\partial\cM}\ne 0$\\{\footnotesize : disallowed}}};


\end{tikzpicture}
\vspace{2mm}
\end{minipage}
\label{fig: u1 cs description 1}
}\hspace{6mm}
\subfigure[$ A_\mu$ and $\phi$ with $ U(1)'$ gauge symmetry]{
\centering
\begin{minipage}[c]{0.28\linewidth}
\centering
\begin{tikzpicture}
\def\length{4}
\def\separation{5}

\draw[thin] (0,0) -- (\length,0) -- (\length,\length) -- (0,\length)--(0,0);

\draw[domain=0:\length,samples=200,smooth, variable=\x,ultra thick] plot ({\x},{1.0+0.1*sin(deg(0.3*\x))+0.2*sin(deg(2*\x))+0.2*sin(deg(3*\x))});

\draw[domain=0:\length,samples=200,smooth, variable=\x,ultra thick] plot ({\x},{3.0+0.1*sin(deg(0.3*\x))+0.2*sin(deg(2*\x))+0.2*sin(deg(3*\x))});

\node[] at ({0.4*\length},1.3) {$A_\mu$};

\node[] at ({0.5*\length},0.2) {$\scriptstyle (\partial_0 -v \partial_1)\lambda|_{\partial\cM}=0${\footnotesize:allowed}};

\draw[-latex] ({0.2*\length},{0.28*\length}) to[bend right] ({0.5*\length},{0.15*\length});
\draw[-latex] ({0.2*\length},1.6) -- ({0.2*\length},3.1); 

\node[] at ({0.56*\length},{0.52*\length}) {\shortstack{$\scriptstyle (\partial_0 -v \partial_1)\lambda|_{\partial\cM}\ne 0$\\{\footnotesize : disallowed}}};

\node[] at ({0.1*\length},{0.52*\length}) {$\phi$};


\end{tikzpicture}
\vspace{2mm}
\end{minipage}
\label{fig: u1 cs description 2}}\hspace{6mm}
\subfigure[$A_\mu,A'_\mu,\cdots$ and $\phi$ with full $U(1)$ gauge symmetry]{
\centering
\begin{minipage}[c]{0.28\linewidth}
\centering
\begin{tikzpicture}
\def\length{4}
\def\separation{5}

\draw[thin] (0,0) -- (\length,0) -- (\length,\length) -- (0,\length)--(0,0);

\draw[domain=0:\length,samples=200,smooth, variable=\x,ultra thick] plot ({\x},{1.0+0.1*sin(deg(0.3*\x))+0.2*sin(deg(2*\x))+0.2*sin(deg(3*\x))});

\draw[domain=0:\length,samples=200,smooth, variable=\x,ultra thick] plot ({\x},{3.0+0.1*sin(deg(0.3*\x))+0.2*sin(deg(2*\x))+0.2*sin(deg(3*\x))});

\node[] at ({0.4*\length},1.3) {$A_\mu$};

\node[] at ({0.5*\length},0.2) {$\scriptstyle (\partial_0 -v \partial_1)\lambda|_{\partial\cM}=0${\footnotesize:allowed}};

\draw[-latex] ({0.2*\length},{0.28*\length}) to[bend right] ({0.5*\length},{0.15*\length});

\node[] at ({0.4*\length},3.3) {$A'_\mu$};

\draw[-latex] ({0.2*\length},1.6) -- ({0.2*\length},3.1); 

\node[] at ({0.56*\length},{0.52*\length}) {\shortstack{$\scriptstyle (\partial_0 -v \partial_1)\lambda|_{\partial\cM}\ne 0$\\{\footnotesize : allowed}}};

\node[] at ({0.1*\length},{0.52*\length}) {$\phi$};

\end{tikzpicture}
\vspace{2mm}
\end{minipage}
\label{fig: u1 cs description 3}}
	\caption{Three equivalent delineations of gauge field configurations and gauge symmetry. \\ (a) $A_\mu, A'_\mu, \cdots $, which are connected by disallowed gauge transformation, are independent physical degrees of freedom with the restricted $U(1)'$ gauge symmetry. (b) Physical gauge configurations $A_\mu,A'_\mu,\cdots$ can be parametrized by the edge mode $\phi$ (would-be gauge mode) and a (reference) gauge configuration $A_\mu$ with the restricted $U(1)'$ gauge symmetry. (c) With full $U(1)$ gauge symmetry, we have $A_\mu,A'_\mu,\cdots$ and the edge mode $\phi$ (Stueckelberg field).}
	\label{fig: u1 cs}
\end{figure}

It is well-known that the Chern-Simons action is not gauge invariant. Together with the boundary term~\eqref{eq: bdy term for cs}, the total action is still not gauge invariant under the gauge transformation of $A\to A+ d\lambda $ in general:
\begin{align}\label{eq:Gauge Transformation of Total Action}
    \delta S_{tot}\,=\, \int_{\partial \cM} d^2x \;  \partial_1\lambda \big[ 2 A_0 -2v  A_1 + \partial_0 \lambda- v\partial_1\lambda\big]\ ,
\end{align}
One way to demand the gauge invariance is to restrict the gauge parameter $\lambda $ such that
\begin{align}
    (\partial_0 -v\partial_1) \lambda\,\big|_{\partial \cM }\,=\,0\ . 
\end{align}
Then, the variation of action under the restricted gauge transformation becomes
\begin{align}
	 \delta S_{tot}\,=\, 2 \int_{\partial \cM} d^2x  \partial_1 \lambda (A_0-vA_1)\, =\, 2 \int_{\partial \cM} d^2x ( \partial_0\lambda - v \partial_1 \lambda )A_1\,=\,0\ .
\end{align}
where we used the equation of motion for the gauge field $A$. In this case, when two configurations $A$ and $A'$ are connected by disallowed ``gauge transformation'' such that 
\begin{align}
    A'_\mu\,=\, A_\mu+ \partial_\mu \lambda\hspace{8mm}\mbox{with}\quad  (\partial_0 -v\partial_1) \lambda\big|_{\partial \cM }\ne 0\ ,
\end{align}
$A'$ and $A$ are not redundant configurations, but physically distinct ones. This implies that there are more physical degrees of freedom which would have been removed by the boundary gauge transformation that is not allowed in this set-up (See Fig.~\ref{fig: u1 cs description 1}). This new physical degree of freedom is called as a would-be gauge mode or an edge mode. This is the common way to deal with the edge mode in the $U(1)$ Chern-Simons theory.

In this description of the edge mode with physically distinct configurations $A_\mu, A'_\mu$ in Fig.~\ref{fig: u1 cs description 1}, there could be conceptual confusion about the boundary condition. By definition, the action is not invariant under the variation of the gauge field along the direction of the disallowed gauge transformation, which might give rise to a misunderstanding that a fluctuation along the direction of the disallowed gauge transformation is frozen. Hence, as a boundary condition of $A_\mu$, one might mistakenly choose one particular configuration $A_\mu$ among all physical configuration $A_\mu,A'_\mu$ that one should have included. In fact, the variation along the disallowed direction does not lead to the boundary condition.\footnote{Hence, one can obtain the correct result without the boundary condition from the variation along the disallowed gauge transformation.} One simple way to clarify this issue is to separate the degree of freedom for edge mode living on the boundary from the bulk gauge field.
\begin{align}
    A(x^0,x^1,x^2)\;\;\longrightarrow\;\; A(x^0,x^1,x^2)+a(x^0,x^1)  \ ,\label{eq: cs bulk edge separation}
\end{align}
where $A$ on the right hand side denotes the ``bulk gauge field'' which is restricted to the boundary condition while $a(x^0,x^1)$ denotes the edge mode  on the boundary. For this, we utilize the boundary ``gauge transformation'' by $\phi(x^0,x^1)$ (\ie \, $a_\mu = \partial_\mu \phi$ in Eq.~\eqref{eq: cs bulk edge separation}), and we promote it as dynamical edge mode.
\begin{align}
    S_{tot}\,=\, S_{CS}+ \int_{\partial \cM} d^2x \;  \big[ (A_1+\partial_1 \phi)(A_0-v A_1 +\partial_0 \phi- v\partial_1\phi)+ \partial_1\phi A_0 -\partial_0\phi A_1 \big]\ .
\end{align}
Then, the path integral measure can be written as
\begin{align}
    \int {\cD A\;\cD \phi\over U(1)'}\ ,
\end{align}
where $U(1)'$ denotes the restricted gauge symmetry of $A$ with $(\partial_0 -v \partial_1) \lambda\big|_{\partial \cM}=0$, and the configuration of the gauge field $A$ is restricted accordingly.

Instead of parameterization of the physical configuration by $A_\mu, A'_\mu,\cdots$, we parametrize them by the edge mode $\phi$ and a specific gauge configuration $A_\mu$ (See Fig.~\ref{fig: u1 cs description 2}). Then the variation along the direction of the disallowed gauge transformation in the previous discussion with $A_\mu,A'_\mu$ can be thought as the variation with respect to the edge mode, which does leads to the boundary condition but the equation of motion of the edge mode. At the same time, one can now impose the boundary condition on $A_\mu$ along that direction to freeze the fluctuation.

It is not easy to deal with the restricted field configuration with the restricted gauge symmetry $U(1)'$. This restriction can easily be removed by extending the gauge field configuration and the gauge symmetry at the same time:
\begin{align}
    A_\mu(x^\mu) \quad&\longrightarrow \quad A_\mu(x^\mu) + \partial_\mu \lambda(x^\mu) \ ,\label{eq: cs full u1 1}\\
    \phi(x^0,x^1)\quad &\longrightarrow \quad \phi(x^0,x^1) - \lambda\big|_{x^2=0}\ , \label{eq: cs full u1 2}
\end{align}
where the gauge parameter $\lambda$ is not restricted and $\phi$ plays a role of Stueckelberg field~\cite{Popov:2021gsa}. Hence, we have the full unrestricted $U(1)$ gauge symmetry, and the gauge field $A$ is unrestricted, too (See Fig.~\ref{fig: u1 cs description 3}). In the description of Fig.~\ref{fig: u1 cs description 3}, the path integral can be written as
\begin{align}
    \int {\cD A\; \cD \phi \over U(1)} e^{iS_{tot}[A,\phi]}
\end{align}
where the total action $S_{tot}[A,\phi]$ is given by
\begin{align}
    S_{tot}\,=\,& \int_{\cM } \big[AdA\big] + \int_{\partial \cM} d^2x \; \big[A_1 (A_0-v A_1 ) \big]+ \int_{\partial \cM} d^2x \; \big[ \partial_1 \phi(\partial_0 \phi- v\partial_1\phi) \big]\cr
    &+ \int_{\partial \cM} d^2x \;  \big[ 2A_0\partial_1\phi -2v A_1 \partial_1\phi \big]\ .
\end{align}
%
%
%
This action can be understood as the action for the chiral boson (edge mode) coupled to the bulk $U(1)$ Chern-Simons theory. 

Now using the full $U(1)$ gauge symmetry~\eqref{eq: cs full u1 1} and \eqref{eq: cs full u1 2}, we will fix the bulk gauge field to obtain the boundary action for the chiral boson coupled to the background field. It is convenient to use different coordinates, $y^\mu$, defined by
\begin{align}
    y^0\,=\, x^0\;\;,\quad y^1\,=\,x^1+vx^0 \;\;,\quad y^2\,=\,x^2\ .
\end{align}
In the $y^\mu$ coordinates, the gauge field can be written as 
\begin{align}
    \tilde{A}_0\,=\, A_0-vA_1\;\;,\quad \tilde{A}_1\,=\, A_1\;\;,\quad \tilde{A}_2\,=\, A_2\ .
\end{align}
Accordingly, the total action becomes
\begin{align}
    S_{tot}\,=\,& \int_{\cM } \;\tilde{A}d\tilde{A}  + \int_{\partial \cM} d^2y \;\big( \tilde{A}_1 \tilde{A}_0  +  \tilde{\partial}_1 \phi\tilde{\partial}_0 \phi  + 2\tilde{\partial}_1\phi \tilde{A}_0\big)\ .\label{eq: cs coupled to edge tilde}
\end{align}
and, the boundary condition becomes
\begin{align}
    \tilde{A}_0\big|_{y^2=0}\,=\, J(y^0,y^1) \;\;:\;\; \mbox{fixed}\ .
\end{align}
This boundary condition allows us to fix the temporal gauge in the bulk to be
\begin{align}
    \tilde{A}_0(y)\,=\, J(y^0,y^1) \hspace{10mm}\mbox{for}\quad y\in \cM\ .
\end{align}
The Gauss constraint $ \tilde{\partial}_1\tilde{A}_2 - \tilde{\partial}_2 \tilde{A}_1=0$ can be solved by
\begin{align}
    \tilde{A}_1\,=\,\tilde{\partial}_1\lambda \;\;,\quad \tilde{A}_2\,=\, \tilde{\partial}_2 \lambda\ .
\end{align}
Using the gauge symmetry, $\lambda$ can be gauged away to fix the $\tilde{A}_1$:
\begin{alignat}{3}
    &\tilde{A}_1\,=\, 0 \quad &&\mbox{for} \quad x^1\in \mathbb{R}\\
    &\tilde{A}_1\,:\,\mbox{constant} \quad &&\mbox{for} \quad x^1\in S^1
\end{alignat}
Then the action~\eqref{eq: cs coupled to edge tilde} (for $x^1\in \mathbb{R}$) becomes
\begin{align}
    S_{tot}\,=\,&  \int_{\partial \cM} d^2y \;\big(  \tilde{\partial}_1 \phi\tilde{\partial}_0 \phi  + 2\tilde{\partial}_1\phi J(y)\big)\ ,\\
    \,=\,& \int_{\partial \cM} d^2x \;\big[  \partial_1 \phi (\partial_0 \phi -v\partial_1\phi  ) + 2\partial_1\phi J(x)\big]\ .
\end{align}
Note that when we introduced the edge mode $\phi$ in Eq.~\eqref{eq: cs bulk edge separation}, there is a trivial redundancy in a constant shift of $\phi$. Hence, one has to impose the equivalence relation on $\phi$:
\begin{align}
    \phi \;\sim \; \phi + c\hspace{10mm}\ ,
\end{align}
where $c$ is a constant. This constant shift gauging eliminates the zero mode of the chiral boson.

One can also obtain the same result from the description with $A_\mu,A'_\mu,\cdots$ in Fig.~\ref{fig: u1 cs description 1} as in the literature. But we find it more convenient to explain the JT gravity and BF theory from the description with $A_\mu$ and $\phi$ in Fig.~\ref{fig: u1 cs description 3}.

\section{Jackiw-Teitelboim Gravity for Asymptotic $AdS_2$}
\label{sec: jt gravity for nearly ads2}

\begin{figure}[t!]
\centering
\subfigure[Variation of metric and dilaton ]{
\begin{minipage}[c]{0.45\linewidth}
\centering
\begin{tikzpicture}
\draw[domain=0:2*pi,samples=200,smooth, variable=\x] plot ({(2.0)*cos(deg(\x))},{(2.0))*sin(deg(\x))});
\draw[domain=0:2*pi,samples=200,smooth, variable=\x,fill=black!5] plot ({(1.8+0.01*(1*cos(deg(5*\x))+2*cos(deg(3*\x))+3*cos(deg(8*\x))+10*cos(deg(11*\x))))*cos(deg(\x))},{(1.8+0.01*(1*cos(deg(5*\x))+2*cos(deg(3*\x))+3*cos(deg(8*\x))+10*cos(deg(11*\x))))*sin(deg(\x))});
\node[] at (0,0.5) {$ g_{\mu\nu}+\delta g_{\mu\nu}$};
\node[] at (0,0) {$ \phi+\delta \phi$};
\node[] at (0,-0.5) {$\partial \cM \text{ : fixed}$};
\end{tikzpicture}
    \vspace{3mm}
\end{minipage}
\label{fig:variation g phi}
}
\subfigure[Variation of wiggling boundary]{
\begin{minipage}[c]{0.45\linewidth}
\centering
\begin{tikzpicture}
\draw[domain=0:2*pi,samples=200,smooth, variable=\x] plot ({(2.0)*cos(deg(\x))},{(2.0))*sin(deg(\x))});
\draw[domain=0:2*pi,samples=200,smooth, variable=\x,fill=black!5] plot ({(1.7+0.01*(1*cos(deg(5*\x))+2*cos(deg(3*\x))+3*cos(deg(8*\x))+10*cos(deg(11*\x))))*cos(deg(\x))},{(1.7+0.01*(1*cos(deg(5*\x))+2*cos(deg(3*\x))+3*cos(deg(8*\x))+10*cos(deg(11*\x))))*sin(deg(\x))});
\draw[domain=0:2*pi,samples=200,smooth, variable=\x,densely dashed] plot ({(1.75+0.01*(1*cos(deg(2*\x))+2*cos(deg(7*\x))+3*cos(deg(12*\x))+10*cos(deg(14*\x))))*cos(deg(\x))},{(1.75+0.01*(1*cos(deg(2*\x))+2*cos(deg(7*\x))+3*cos(deg(12*\x))+10*cos(deg(14*\x))))*sin(deg(\x))});
\node[] at (0,0.5) {$g_{\mu\nu}\text{ : fixed}$};
\node[] at (0,0) {$\phi\text{ : fixed}$};
\node[] at (0,-0.5) {$\partial\cM+\delta \partial\cM$};
\end{tikzpicture}
    \vspace{3mm}
\end{minipage}
\label{fig:variation boundary}
}
	\caption{The variation of the JT action can be split into two part. (a) The variation of the metric and the dilaton with the wiggling boundary fixed leads to the boundary condition for the bulk metric and the bulk dilaton. (b) The variation of the wiggling boundary with the metric and the dilaton fixed gives the equation of motion of the edge mode.}
	\label{fig: variation}
\end{figure}
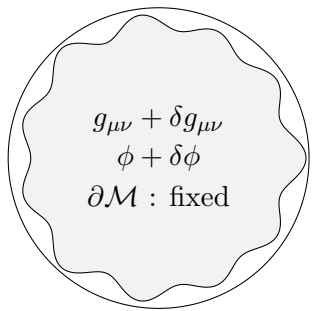
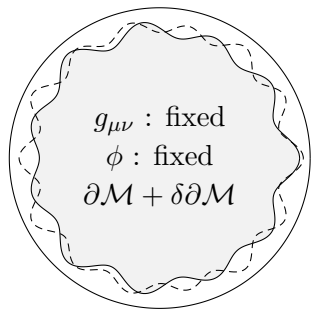 
The action\footnote{The overall sign is opposite to that of the literature for ``nearly-AdS''~\cite{Maldacena:2016upp,Saad:2019lba}. Here, we determine the overall sign for the positive dilaton and the stability of the Schwarzian action. We will discuss the dimensional reduction point of view in Section~\ref{sec: conclusion}.} of Jackiw-Teitelboim gravity in a Euclidean manifold $\cM$ with boundary $\partial \cM$ is given by
\begin{align}
    I_{\rm JT}\,=\,  \int_{\cM} d^2x\, \sqrt{g}\, \phi\, (R+2) + 2 \int_{\partial \cM} du\, \sqrt{h} \, \phi\, \big(K - K_0\big)\ , \label{eq: jt action}
\end{align}
where $\phi$ and $g_{\mu\nu}$ are the dilaton and the bulk metric. The boundary $\partial \cM$ is parameterized by the coordinate $u$\,, and $h\equiv h_{uu}$ is the induced metric on the boundary. $K$ is the extrinsic curvature evaluated with the metric $g_{\mu\nu}$, whereas $K_0$ is the same quantity evaluated with the background metric $g_{0\,\mu\nu}$. Hence, $K_0$ plays a role of a counter-term which makes the free energy of the background vanish. In this work, for a given background geometry, we will consider geometries which do not change the global properties or topology of the background, which will correspond to the smooth gauge transformation in contrast to the large gauge transformation. Hence, from the beginning of the formulation we choose the background that we are interested in, and the global information of the background is incorporated into $K_0$.

We will study the gravitational edge mode in the JT gravity as a wiggling boundary in Section~\ref{sec: wiggling boundary} as in \cite{Maldacena:2016upp} and as a would-be gauge mode in Section~\ref{sec: stueckelberg field} as in \cite{Carlip:2005tz}. These two descriptions are, in fact, equivalent, and we will explain the relation explicitly.

\subsection{Description 1:  Wiggling Boundary}
\label{sec: wiggling boundary}

In this section, we revisit the derivation of the Schwarzian action for the Jackiw-Teitelboim gravity with wiggling boundary. The relation to the edge mode to the wiggling boundary will be clarified in the next section. Let us begin with the variation of the action:
\begin{align}
    \delta I_{JT}\,=\, & \mbox{(bulk E.o.M)}+  \int_{\partial \cM }du \, \sqrt{h}\, \bigg[ \big(\partial_n \phi -  K_0 \phi  \big)  h^{\mu\nu}\delta h_{\mu\nu} +  2 \big(K-K_0\big)\delta \phi \bigg]\cr
    &+2\int_{\delta \partial \cM} du \,\sqrt{h} \, \phi\, \big(K - K_0\big) \ . \label{eq: variation of jt action}
\end{align}
The first line represents the variation of the (bulk) metric and the (bulk) dilaton with the wiggling boundary fixed~(Fig.~\ref{fig:variation g phi}) while the second line corresponds to the variation of the wiggling boundary with the metric and the dilaton fixed~(Fig.~\ref{fig:variation boundary}). For the variation of the first line of Eq.~\eqref{eq: variation of jt action}, we impose the boundary condition that the boundary metric and the boundary dilaton is fixed for a given wiggling boundary. \ie
\begin{align}
    \delta h_{uu}\,=\,\delta \phi \,=\, 0 \hspace{10mm} \mbox{for a fixed wiggling boundary} \ .
\end{align}

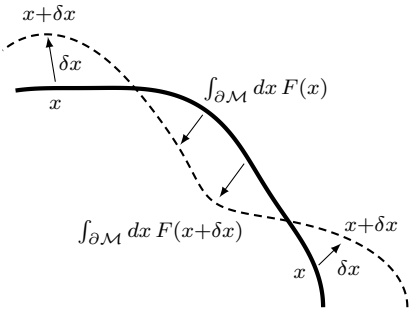
\begin{figure}[t!]
\centering
\begin{tikzpicture}
\draw[domain=0:0.6*pi,samples=200,smooth, variable=\x,ultra thick] plot ({(3+0.01*(1*cos(deg(5*\x))+2*cos(deg(2.5*\x))+3*cos(deg(4*\x))+10*cos(deg(6*\x))))*cos(deg(\x))},{(3+0.01*(1*cos(deg(5*\x))+2*cos(deg(2.5*\x))+3*cos(deg(4*\x))+10*cos(deg(6*\x))))*sin(deg(\x))});
\draw[domain=0:0.6*pi,samples=200,smooth, variable=\x,densely dashed, thick] plot ({(3+0.08*(1*cos(deg(2*\x))+2*cos(deg(7*\x))+3*cos(deg(5*\x))+10*cos(deg(3.5*\x))))*cos(deg(\x))},{(3+0.08*(1*cos(deg(2*\x))+2*cos(deg(7*\x))+3*cos(deg(5*\x))+10*cos(deg(3.5*\x))))*sin(deg(\x))});

\draw[-latex] (1.56,2.56) to (1.26,2.16);

\draw[-latex] ({1.56+0.55},{2.56-0.65}) to ({1.2+0.58},{2.16-0.7});

\draw[-latex] (-0.4,3) to (-0.5,3.6);

\draw[-latex] (3.1,0.56) to (3.4,0.85);

\node[] at (2.85,0.45) {$\scriptstyle x$};
\node[] at (3.5,0.52) {$\scriptstyle\delta x$};
\node[] at (3.8,1.1) {$\scriptstyle x+\delta x$};

\node[] at (-0.4,2.7) {$\scriptstyle x$};
\node[] at (-0.2,3.25) {$\scriptstyle\delta x$};
\node[] at (-0.48,3.9) {$\scriptstyle x+\delta x$};

\node[] at (2.4,2.9) {$\scriptstyle \int_{\partial \cM} dx\, F(x)$};
\node[] at (1.0,1.0) {$\scriptstyle \int_{\partial \cM} dx\, F(x+\delta x) $};

\end{tikzpicture}
	\caption{The variation with respect to the wiggling boundary with the metric and dilaton fixed can be evaluated by the variation of the integrand with respect to the coordinates. }
	\label{fig: variation of boundary detail}
\end{figure}

On the other hands, the variation of the wiggling boundary in the second line of Eq.~\eqref{eq: variation of jt action} can be written as a variation of the integrand with respect to $(r,\theta)$ denoting the coordinates of the $AdS_2$ with the metric and dilaton functions fixed (See Fig.~\ref{fig: variation of boundary detail}). 
\begin{align}
    2\int_{ \partial \cM} du \,\delta_{r,\theta}\bigg[\sqrt{h} \, \phi\, \big(K - K_0\big)\bigg]_{g,\phi:\text{ fixed}}\ .
\end{align}
Unlike the boundary condition for the (bulk) metric and dilaton, this variation gives the equation of motion of the edge mode, \ie,\; Schwarzian mode. This is analogous to the issue on the variation of the $U(1)$ Chern-Simons theory with the description in Fig.~\ref{fig: u1 cs description 1}.

By using the diffeomorphism, one can fix the metric and the dilaton to be~\cite{Witten:2020ert} 
\begin{align}
    ds^2\,=\,& {dr^2\over r^2-r_h^2} + (r^2-r_h^2)\, d\theta^2\ ,\\
    \phi \,=\,& \, r\ ,
\end{align}
where $r_h$ is the location of the tip of the Euclidean black hole. The Euclidean time $\theta$ is periodic with period $\beta$, and the smoothness condition around $r=r_h$ gives
\begin{align}
    r_h\,=\, {2\pi \over \beta}\ .
\end{align}
At the cost of trivializing the bulk geometry and the dilaton profile, the non-trivial information is encoded only in the shape of the boundary. The wiggling boundary curve can be parametrized by $(r(u), \theta(u))$ where $u\in [0,\beta]$. 
\begin{figure}[t!]
\centering
\begin{tikzpicture}
\draw[domain=0:2*pi,samples=200,smooth, variable=\x] plot ({(2.0)*cos(deg(\x))},{(2.0))*sin(deg(\x))});
\draw[domain=0:2*pi,samples=200,smooth, variable=\x,fill=black!5] plot ({(1.8)*cos(deg(\x))},{(1.8))*sin(deg(\x))});

\draw[domain=0:2*pi,samples=200,smooth, variable=\x] plot ({6.8+(2.0)*cos(deg(\x))},{(2.0))*sin(deg(\x))});
\draw[domain=0:2*pi,samples=200,smooth, variable=\x,fill=black!5] plot ({6.8+(1.8+0.01*(1*cos(deg(5*\x))+2*cos(deg(9*\x))+3*cos(deg(18*\x))+4*cos(deg(13*\x))))*cos(deg(\x))},{(1.8+0.01*(1*cos(deg(5*\x))+2*cos(deg(9*\x))+3*cos(deg(18*\x))+4*cos(deg(13*\x))))*sin(deg(\x))});

\node[circle,fill=black,scale=0.5, label=below:{$(\rho,u)$}] (n1) at (1.0,0.6) {};

\node[circle,fill=black,scale=0.5, label=below:{$(r,\theta)$}] (n2) at (7.5,0.8) {};

\draw[-latex] (n1) to[bend left]node[midway, sloped, above]{$\big(r(\rho,u),\theta(\rho,u)\big)$} (n2);
\node[] at (0,-2.5) {Base AdS$_2$};
\node[] at (6.8,-2.5) {Target AdS$_2$};

\end{tikzpicture}
	\caption{The wiggling boundary can be understood as a map from $S^1$ to the wiggling boundary of AdS$_2$. It is convenient to extend the circle $S^1$ to the exact AdS$_2$ to define \textit{base AdS$_2$}. We can consider a map from the base AdS$_2$ to the target AdS$_2$.}
	\label{fig:wiggling boundary}
\end{figure}
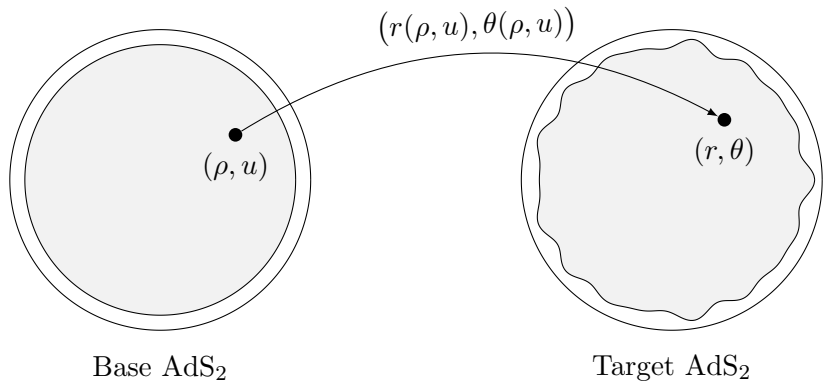 
One can extend this parametrization of the boundary curve to the AdS$_2$. Namely, let us consider the exact EAdS$_2$ without wiggling boundary, 
\begin{align}
    ds^2\,=\, {d\rho^2\over \rho^2-r_h^2} + (\rho^2-r_h^2) du^2 \ .\label{eq: base ads metric}
\end{align}
The boundary surface is parametrized by
\begin{align}
    \rho\,=\, {1\over \epsilon}\ .
\end{align}
We consider a map from the \textit{base AdS} $(\rho,u)$ to the target $AdS$ $(r,\theta)$ such that the constant $\rho=\epsilon^{-1}$ surface is mapped to the wiggling boundary in $(r,\theta)$ space
\begin{align}
    r\,=\, &r\big(\epsilon^{-1}, u)\,\equiv\,r(u)\ ,\\
    \theta\,=\,& \theta\big(\epsilon^{-1}, u)\,\equiv\, \theta(u)\ ,
\end{align}
and the boundary metric is identified
\begin{align}
    {1\over \epsilon^2}-r_h^2\,=\, {r'^2\over r^2-r_h^2 } + (r^2-r_h^2)\theta'^2 \ .\label{eq: boundary metric}
\end{align}
From the boundary metric matching condition~\eqref{eq: boundary metric}, one can express $r(u)$ in terms of $\theta(u)$ perturbatively.
\begin{align}
    r\,=\,{1\over \epsilon \theta'}+ \epsilon\bigg( -{\theta''^2\over 2\theta'^3}+{1\over 2}r_h^2 \theta' -{r_h^2\over 2\theta'} \bigg)+\mathcal{O}(\epsilon^3)\ ,\label{eq: wiggling r sol}
\end{align}
where the overall sign on the right hand side is chosen in a way that $\theta(u)$ is the increasing function of $u$. Then, the extrinsic curvature can be evaluated to be
\begin{align}
    K\,=\,& {\big( \mathfrak{t} , \nabla_{\mathfrak{t}} \mathfrak{n} \big)\over \big( \mathfrak{t}, \mathfrak{t} \big)} \,=\,{  {6r\over r^2-r_h^2}r'^2\theta'+  2r(r^2-r_h^2)\theta'^3-2r''\theta'+2r'\theta'' \over 2\bigg[ {r'^2\over r^2-r_h^2}+ (r^2-r_h^2) \theta'^2 \bigg]^{3\over 2}}\ ,
\end{align}
where we chose the tangent vector $\mathfrak{t}$ and the (outward) normal vector $\mathfrak{n}$ to be
\begin{align}
    \mathfrak{t}\,=\, (r',\theta') \;\;,\quad \mathfrak{n}\,=\, {1\over \sqrt{r'^2\over (r^2-r_h^2)}+(r^2-r_h^2)\theta'^2} \bigg((r^2-r_h^2)\theta', -{r'\over r^2-r_h^2}\bigg) \ .
\end{align}
From Eq.~\eqref{eq: wiggling r sol}, the extrinsic curvature can be expanded with respect to $\epsilon$:
\begin{align}
    K\,=\,&  1 +\epsilon^2 \bigg[{\theta'''\over \theta'}-{3\over 2}{\theta''^2\over \theta'^2}+{1\over 2}r_h^2 \theta'^2 \bigg]+\mathcal{O}(\epsilon^4)\ .
\end{align}
And we chose the counter term $K_0$ in the boundary action~\eqref{eq: jt action} to be the extrinsic curvature of the geometry without wiggling (\ie \; $\theta(u)=u$): 
\begin{align}
    K_0\,=\, 1 + {1\over 2}\epsilon^2 r_h^2 +\cO(\epsilon^4)\,.\label{eq: counter term}
\end{align}
{This counter-term is different from that in the previous literature. It turns out that this new counter-term plays an important role in the derivation of the Schwarzian action. In the action~\eqref{eq: jt action}, the bulk part vanishes while the boundary action becomes
\begin{align}
    I_{JT}\,=\,&  2 \int du \sqrt{{1\over\epsilon^2}-r_h^2 }  \bigg[{1\over \epsilon \theta'}+ \epsilon\bigg( -{\theta''^2\over 2\theta'^3}+{1\over 2}r_h^2 \theta' -{r_h^2\over 2\theta'} \bigg)+\cdots \bigg]\cr
    &\times \epsilon^2\bigg[ {\theta'''\over \theta'}-{3\over 2}{\theta''^2\over \theta'^2}+{1\over 2}r_h^2 \theta'^2 -{1\over 2}r_h^2 \bigg]\ ,\\
    \,=\,& -2 \int du \, \bigg[ -{1\over \theta'}Sch[\theta,u] -{1\over 2}r_h^2 \theta' +{1\over 2\theta'}r_h^2\bigg]\ .\label{eq: jt boundary action wiggling}
\end{align}
where the Schwarzian derivative $Sch[\theta,u]$ is defined by
\begin{align}
    Sch[\theta,u]\,\equiv \, {\theta'''\over \theta'}-{3\over 2}{\theta''^2\over \theta'^2} \ .
\end{align}
First, note that $\theta'^{-1}$ in front of the Schwarzian derivative comes from the dilaton solution $\phi\,=\, r$. This is one of the main differences from the previous derivation of the Schwarzian action. The bulk dilaton field is the bulk Lagrangian multiplier which is determined by the bulk equation of motion. And we choose the boundary value of the dilaton which is consistent with the dilaton solution.

The second term in Eq.~\eqref{eq: jt boundary action wiggling} gives a constant $r_h^2\beta={2\pi  r_h}$ to the action. The last term in~\eqref{eq: jt boundary action wiggling} comes from the term of order $\cO(\epsilon^2)$ in the counter term $K_0$~\eqref{eq: counter term}. Now using the inversion formula for the Schwarzian derivative
\begin{align}
    Sch[\theta,u]\,=\, - \theta'^2Sch[u,\theta]\ ,\label{eq: inversion formula of schwarzian derivative}
\end{align}
we have
\begin{align}
    I_{JT}  \,=\, & r_h^2 \int d \theta  - 2 \int d\theta \, \bigg[ Sch[u,\theta] + {1\over 2}r_h^2 u'^2 \bigg]\ .\label{eq: schwarzian action wiggling}
\end{align}
Note that the last term comes from the counter term. This action vanishes for the trivial map $u=\theta$ which is expected by the choice of the counter term. Also note that the sign in front of the Schwarzian derivatives is important for the stability of the Euclidean action. In the semi-classical analysis around the classical solution $u(\theta)=\theta$, the quadratic action of the quantum fluctuation $\epsilon$ is bounded below with the minus sign in front of the Schwarzian derivative.
\begin{align}
    u\,=\, \theta + {\beta\over 2\pi } \sum_n \epsilon_n\, e^{{2\pi i n \over \beta} \theta} \;\;\to \;\; S_{AdS}^{\text{\tiny edge}}\,=\,  {4\pi^2\over \beta}\sum_{n\in \mathbb{Z}/ \{-1,0,1\}}n^2(n^2-1)\epsilon_{-n}\epsilon_n + \cO(\epsilon^3)\ .
\end{align}

In deriving the Schwarzian action, we had to invert $\theta(u)$ into $u(\theta)$. This inversion naturally leads\footnote{At this moment, we assume that the path integral measure is trivial with $\theta(u)$.} to the well-known path integral measure of the Schwarzian action:
\begin{align}
    \prod_{u} \cD \theta(u) \,=\, \prod_{\theta } {\cD u(\theta)\over u'}\ . 
\end{align}

Now, we will explain how the $sl(2,\mathbb{R})$ gauging appears in our formulation. The path integral of the JT gravity is reduced to the path integral of the map from the base AdS$_2$ to the target AdS$_2$\footnote{At this moment, the introduction of the base AdS$_2$ and the map from the base AdS$_2$ to the target AdS$_2$ would not look natural although one can easily explain how the $PSL(2,\mathbb{R})$ gauging appears. In Section~\ref{sec: stueckelberg field}, we will clarify the origin of the base AdS$_2$ and the map. }
\begin{align}
    (\rho,u) \;\;\longrightarrow \;\; (r,\theta)\ .
\end{align}
However, the base AdS has the $sl(2,\mathbb{R})$ isometry, and the parametrization of the target AdS cannot distinguish the isometry of the base AdS. This leads to the redundancy in the parametrization of the target Ads which should be gauged to avoid overcounting. For example, the isometry of the Poincare coordinates 
\begin{align}
    ds^2\,=\, {dt^2 + dz^2\over z^2}\ ,
\end{align}
is given by
\begin{align}
    t'\,=\,&{bd + (ad+bc)t +ac (t^2+z^2) \over (ct + d)^2+c^2 z^2}\ ,\\
    z'\,=\,&{z\over (ct +d)^2+c^2z^2} \ ,
\end{align}
where $a, b, c, d$ are constants with $ad-bc=1$. And on the boundary $t'$ becomes
\begin{align}
    t'\big|_{z=0}\,=\,\lim_{z\to 0}{bd + (ad+bc)t +ac (t^2+z^2) \over (ct + d)^2+c^2 z^2}\,=\, {at + b\over ct +d} \ ,
\end{align}
which is well-known $sl(2,\mathbb{R})$ transformation of the boundary coordinate $t$.

Using the coordinate transformation from the Poincare coordinates $(z,t)$ to our (base AdS) coordinates $(\rho,u)$
\begin{align}
    z\,=\,{r_h\over \rho+ \sqrt{\rho^2-r_h^2}\cos{r_h u}}\ ,\\
    t\,=\, {\sqrt{\rho^2-r_h^2}\sin {r_h u}\over \rho+ \sqrt{\rho^2-r_h^2}\cos{r_h u}}\ .
\end{align}
The time on the boundary $z=0$ in the Poincare coordinates is identified with the boundary time $u$ as follows.
\begin{align}
    t|_{r\to \infty}\,=\, {\sin r_hu\over 1+\cos r_h u}\,=\, \tan{r_hu\over 2} \ .
\end{align}
Therefore, under the isometry of the base AdS, the coordinate $u$ on the boundary is transformed as
\begin{align}
    \tan{r_h u\over 2}\quad\longrightarrow \quad {a  \tan{r_h u \over 2}+b \over c  \tan{r_h u\over 2} +d }\ .\label{eq: sl2 gauge isometry}
\end{align}
Note that the Schwarzian action in~\eqref{eq: schwarzian action wiggling} is invariant under the $PSL(2,\mathbb{R})$ transformation of $u$~\eqref{eq: sl2 gauge isometry} which comes from the isometry of the base AdS. And this $PSL(2,\mathbb{R})$ should be gauged because of the redundant description of the target AdS. It is important to note that the isometry of the target AdS, which can change the dilaton solution, should not be gauged in the Schwarzian action.

In the next section, we will derive the Schwarzian action from the would-be gauge mode and will discuss the relation to the wiggling boundary.

\subsection{Description 2: Would-be Gauge Mode}
\label{sec: stueckelberg field}

Let us begin with a Euclidean AdS$_2$ without the wiggling boundary. Namely, for the coordinates $X^\mu\,=\,(\rho,u)$, the boundary of the EAdS$_2$ is the surface of constant $\rho$. We consider the Fefferham-Graham gauge given by
\begin{align}
    ds^2\,=\,{d\rho^2\over \rho^2-r_h^2} + G_{uu}(\rho,u) du^2\ ,\label{eq: fg gauge x}
\end{align}
where $r_h$ is a constant corresponding to the radius of the black hole that we are interested as a background. The boundary metric is expanded as

\begin{align}
   G_{uu}(\rho,u)\,=\, \rho^2 G^{(0)}(u)  + G^{(2)}(u) +\cdots \ .
\end{align}

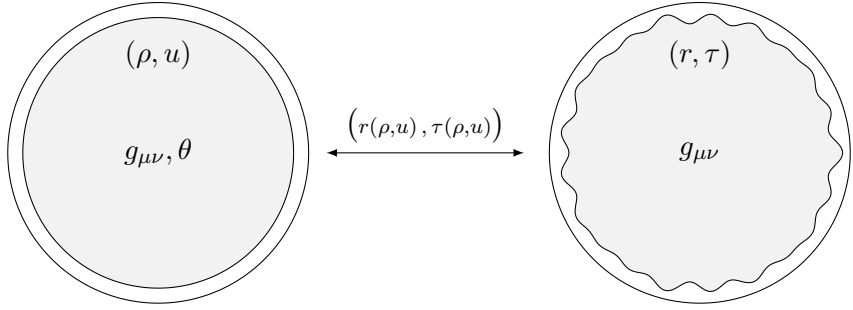
\begin{figure}[t!]
\centering
\begin{tikzpicture}



\draw[domain=0:2*pi,samples=200,smooth, variable=\x] plot ({(2.0)*cos(deg(\x))},{6
+(2.0)*sin(deg(\x))});
\draw[domain=0:2*pi,samples=200,smooth, variable=\x,fill=black!5] plot ({(1.8)*cos(deg(\x))},{6+(1.8)*sin(deg(\x))});

\draw[domain=0:2*pi,samples=200,smooth, variable=\x] plot ({7.2+(2.0)*cos(deg(\x))},{6+((2.0))*sin(deg(\x))});
\draw[domain=0:2*pi,samples=200,smooth, variable=\x,fill=black!5] plot ({7.2+(1.8+0.01*(1*cos(deg(7*\x))+2*cos(deg(10*\x))+3*cos(deg(17*\x))+4*cos(deg(21*\x))))*cos(deg(\x))},{6+(1.8+0.01*(1*cos(deg(7*\x))+2*cos(deg(10*\x))+3*cos(deg(17*\x))+4*cos(deg(21*\x))))*sin(deg(\x))});

\node[] (n1) at (2.1,6) {};
\node[] (n2) at (5.0,6) {};

\draw[latex-latex] (n1) to node[midway, sloped, above]{$\scriptstyle \big(r(\rho,u)\,,\,\tau(\rho,u)\big)$} (n2);

\node[] at (7.2,7.3) {$(r,\tau)$};
\node[] at (0,7.3) {$(\rho,u)$};

\node[] at (7.2,6) {$g_{\mu\nu}$};
\node[] at (0,6) {$g_{\mu\nu}, \theta$};


\end{tikzpicture}
	\caption{The would-be gauge mode is ``eaten'' by the wiggling boundary via the radial and boundary diffeomorphism. }
	\label{fig:would-be boundary}
\end{figure}
One can consider a reference AdS$_2$ geometry with the coordinates $Y^\mu=(r,\tau)$ which is mapped to $X^\mu\,=\,(\rho,u)$ by radial and boundary diffeomorphism (See Fig.~\ref{fig:would-be boundary}):
\begin{align}
    r\,=\, &w(u)\rho  + {1\over \rho} f^{(2)}(u)+\cO\big(\rho^{-3}\big) \label{eq: diffeomorphism 1}\ ,\\
    \tau\,=\,& \theta(u) + {h^{(2)}(u)\over \rho^2 g^{(0)}(u)} +\cO\big(\rho^{-4}\big)\ ,\label{eq: diffeomorphism 2}
\end{align}
Then, while the boundary of the $(\rho,u)$ is the surface of constant $\rho$, the boundary of the $(r,\tau)$ is wiggling by the radial diffeomorphism~\eqref{eq: diffeomorphism 1}. One can still demand that the metric $g_{\mu\nu}$ of the reference AdS $Y^\mu=(r,\tau)$ preserves the Fefferham-Graham gauge~\eqref{eq: fg gauge x}
\begin{align}
    ds^2\,=\,{dr^2\over r^2-r_h^2} + g_{\tau\tau}(r,\tau) d\tau^2\ ,\label{eq: fg gauge y}
\end{align}
and that the expansion of the boundary metric $g_{\tau\tau}$ is truncated.
\begin{align}
   g_{\tau\tau}(r,\tau)\,=\, r^2 g^{(0)}(\tau)  + g^{(2)}(\tau) \ .
\end{align}
%
%
The condition for the Fefferham-Graham gauge~\eqref{eq: fg gauge x} and \eqref{eq: fg gauge y} determines the function $f^{(2)}(u)$ and $h^{(2)}(u)$ in Eq.~\eqref{eq: diffeomorphism 1} and Eq.~\eqref{eq: diffeomorphism 2}
\begin{align}
    f^{(2)}(u)\,=\,& {w'^2\over 4g^{(0)}[\theta(u)] w^3 \theta'^2}-{1\over 4}r_h^2 w +{r_h^2\over 4 w}\ ,\label{eq: fg gauge sol1}\\
    h^{(2)}(u)\,=\,& {w'\over 2w^3 \theta'}\ . \label{eq: fg gauge sol2}
\end{align}
Then the boundary metric $G_{uu}$ can be expressed in terms of $g_{uu}, w(u)$ and $\theta(u)$:
\begin{align}
    &G_{uu}\,=\, g^{(0)}[\theta(u)] w^2 \theta'^2 \rho^2 \cr
    +&  \bigg(g^{(2)}[\theta(u)]+{1\over 2}r_h^2(1-w^2) g^{(0)}[\theta(u)]\bigg)\theta'^2 +{w''\over w} -{3 w'^2\over 2 w^2} - {w' \theta''\over w \theta'} -{g^{(0)\prime}[\theta(u)]\theta'w'\over 2g^{(0)}[\theta(u)]w} \cr
    &+ \mathcal{O}\big(\rho^{-2}\big)\ .
\end{align}
Now in addition to the Fefferman-Graham gauge~\eqref{eq: fg gauge x} and \eqref{eq: fg gauge y}, we further impose the asympotic AdS condition. \ie
\begin{align}
    G^{(0)}(u)\,=\,  1 \ , \label{eq: asymptotic ads condition metric y}
\end{align}
and we also demand that the diffeomorphism in Eqs.~\eqref{eq: diffeomorphism 1} and \eqref{eq: diffeomorphism 1} preserve the asymptotic AdS condition. 
\begin{align}
    g^{(0)}(\tau)\,=\, 1 \ , \label{eq: asymptotic ads condition metric x}
\end{align}
Then, the asymptotic AdS condition gives
\begin{align}
    w\theta' \,=\, 1\ .\label{eq: asymptotic ads condition sol metric}
\end{align}
And one can express the boundary metric in terms of $\theta$
\begin{align}
    G_{uu}\,=\, \rho^2 +  \bigg(\theta'^2 g^{(2)}[\theta(u)]+{1\over 2}r_h^2(\theta'^2-1) \bigg) -{\theta'''\over \theta'} + {3 \theta''^2\over 2 \theta'^2} + \mathcal{O}\big(\rho^{-2}\big)\ .\label{eq: boundary metric sol}
\end{align}
Now we go back to the JT action in the $X^\mu=(\rho,u)$ coordinate space with the metric $G_{\mu\nu}$:
\begin{align}
    I_{JT}\,=\,&  \int_{\cM} d^2x \sqrt{G} \phi (R[G]+2) + 2 \int_{\partial \cM} du\, \sqrt{H_{uu}}\, \phi \big(K[G] -K_0\big)\ .
\end{align}
Recall that the boundary of $X^\mu=(\rho,u)$ is the surface of constant $\rho$. Hence, the induced boundary metric $H_{uu}$ is the same as $G_{uu}$. After choosing the Fefferman-Graham gauge~\eqref{eq: fg gauge x} and the asymptotic AdS condition~\eqref{eq: asymptotic ads condition metric x}, one can still have residual diffeomorphism~\eqref{eq: diffeomorphism 1} and \eqref{eq: diffeomorphism 2} with Eq.~\eqref{eq: fg gauge sol1}, Eq.~\eqref{eq: fg gauge sol2} and Eq.~\eqref{eq: asymptotic ads condition sol metric} which is, therefore, parametrized by $\theta(u)$. This diffeomorphism is broken on the boundary of AdS, and therefore the would-be gauge mode by this diffeomorphism becomes the gravitational edge mode as in the $U(1)$ Chern-Simons theory with a boundary. As in Section~\ref{sec: u1 cs theory}, we promote the function $\theta(\tau)$ parameterizing the broken diffeomorphism to be a dynamical edge mode. Inserting the metric relation~\eqref{eq: boundary metric sol} into the action, we have the JT action of the metric $g$ coupled to the gravitational edge mode $\theta$:
\begin{align}
    I_{JT}\,=\, & \int_{\cM} d^2x \sqrt{g}\,   \bigg|{\partial Y^\mu\over \partial X^\nu}\bigg|\, \phi\, 
    \big(R\big[G[g,\theta]\big]+2\big) \cr
    &+ 2 \int_{\partial \cM} du\, \sqrt{G_{uu}[g,\theta]}\, \phi \,\big(K\big[G[g,\theta]\big] -K_0\big)\ .\label{eq: jt action with edge mode}
\end{align}
where the counter term $K_0$ is chosen to be the extrinsic curvature of the background geometry in Eq.~\eqref{eq: counter term}. By the definition of the edge mode $\theta$, the action~\eqref{eq: jt action with edge mode} is invariant under the following transformation which corresponds to the restored (residual) broken gauge symmetry by introducing the edge mode.
\begin{align}
    \rho\;&\longrightarrow \; {\rho \over \lambda'(u)}+ \cO(\rho^{-1})\ ,\label{eq: restored sym 1}\\
    u \; &\longrightarrow \; \lambda(u)+\cO(\rho^{-2})\ ,\label{eq: restored sym 2}\\
    \theta\;&\longrightarrow \; \theta\circ \lambda^{-1}\ .\label{eq: restored sym 3}
\end{align}
Using this gauge symmetry \eqref{eq: restored sym 1}$\sim$\eqref{eq: restored sym 3}, one can fix the boundary metric $g^{(2)}$ to be a constant, and the constant value of $g^{(2)}$ is determined to have the smooth geometry without conical defect.
\begin{align}
    g^{(2)}\,=\,&-r_h^2\ . \label{eq: boundary metric g2}
\end{align}
In addition, the (bulk) equation of motion for the dilaton is simpler in the $(r,\tau)$ space where we can determine the dilaton solution to be
\begin{align}
    \phi\,=\, r\,=\, {1\over \theta'}\rho + {1\over \rho}\bigg[{\theta''^2\over 4  \theta'^3}-{1\over 4}r_h^2 \theta +{r_h^2\over 4 \theta}\bigg]+\cO(\rho^{-3})\ .
\end{align}
Inserting them into the action~\eqref{eq: jt action with edge mode}, the bulk action vanishes while the boundary action becomes
\begin{align}
    I_{JT}[\theta]\,=\, & 2\int_{\partial \cM} du \sqrt{G_{uu}[g^{\text{\tiny AdS}},\theta]}\, \phi\, \big(K\big[G[g^{\text{\tiny AdS}},\theta]\big] -K_0\big) \ ,\cr
    \,=\,& -2\int du \;\bigg[ - {1\over \theta'}Sch\big[\theta,u\big] -{1\over 2}r_h^2 \theta' +{1\over 2\theta'}r_h^2 \bigg]\ .
\end{align}
As before, we can invert $\theta(u)$ into $u(\theta)$ to get the Schwarzian action
\begin{align}
    I_{JT}[\theta]\,=\,& r_h^2 \int d\theta  -2\int d\theta \;\bigg[ Sch\big[u,\theta\big] + {1\over 2}r_h^2 u'^2 \bigg]\ .
\end{align}
The path integral measure for the Schwarzian action is induced by the inversion.
\begin{align}
    \prod_{u} \cD \theta(u) \,=\, \prod_{\theta } {\cD u(\theta)\over u'}\ . 
\end{align}
After we fix the boundary metric $g^{(2)}$~\eqref{eq: boundary metric g2} by using the gauge symmetry~\eqref{eq: restored sym 1}$\sim$\eqref{eq: restored sym 3}, there is still the residual gauge symmetry in which $g^{(2)}$ is invariant, which exactly corresponds to the isometry transformation of the background.
\begin{align}
    \tan{r_h u\over 2}\quad\longrightarrow \quad {a  \tan{r_h u \over 2}+b \over c  \tan{r_h u\over 2} +d }\hspace{10mm}\mbox{where}\quad ad-bc\,=\,1 \ . \label{eq: sl2 gauge for wouldbe}
\end{align}
To see this, one can consider the gauge transformation of $g^{(2)}$ induced by the coordinate transformation in \eqref{eq: restored sym 1} and \eqref{eq: restored sym 2}.
\begin{align}
    g^{(2)}=-r_h^2\;\;\longrightarrow \;\; - {1\over 2}r_h^2 -{1\over 2}r_h^2\lambda'^2  -{\lambda'''\over \lambda'} + {3 \lambda''^2\over 2 \lambda'^2}\,=\, -{1\over 2}r_h^2 - Sch\bigg[\tan{r_h \lambda\over 2},u\bigg]\ .
\end{align}
Therefore, the function $\lambda$ which does not change $g^{(2)}=-r_h^2$ is given by
\begin{align}
    \lambda(u)\,=\, {2\over r_h} \arctan\bigg[ {a \tan{r_h u \over 2}+ b\over c \tan{r_h u \over 2}+ d}\bigg] \hspace{10mm}\mbox{where}\quad ad-bc\,=\,1 \ ,
\end{align}
and this leads to the $PSL(2,\mathbb{R})$ gauge symmetry~\eqref{eq: sl2 gauge for wouldbe} of the Schwarzian theory.

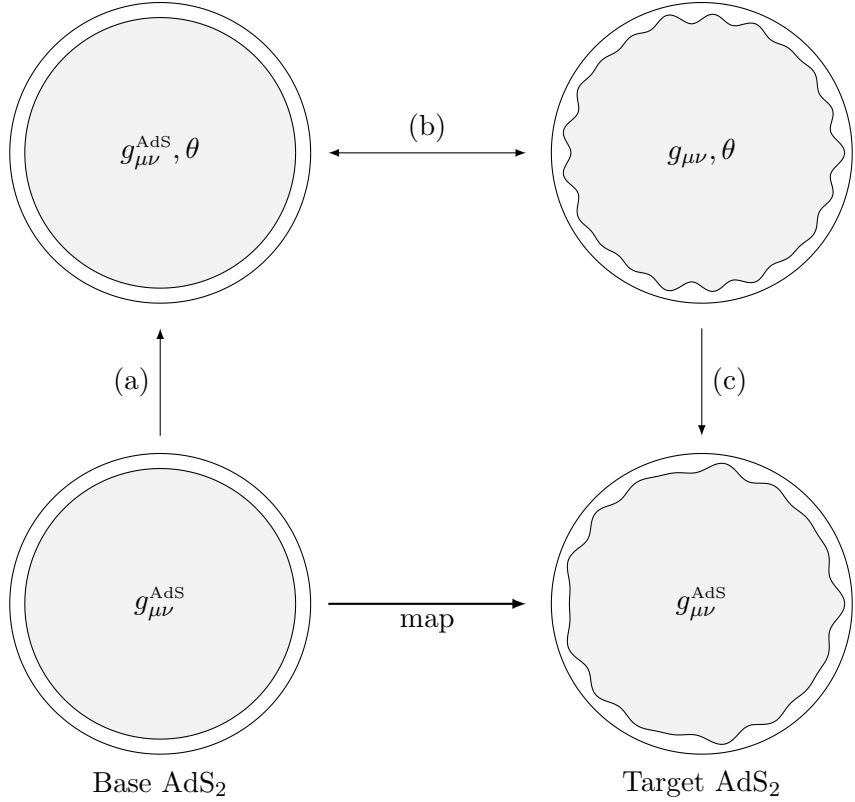
\begin{figure}[t!]
\centering
\begin{tikzpicture}
\draw[domain=0:2*pi,samples=200,smooth, variable=\x] plot ({(2.0)*cos(deg(\x))},{(2.0))*sin(deg(\x))});
\draw[domain=0:2*pi,samples=200,smooth, variable=\x,fill=black!5] plot ({(1.8)*cos(deg(\x))},{(1.8))*sin(deg(\x))});

\draw[domain=0:2*pi,samples=200,smooth, variable=\x] plot ({7.2+(2.0)*cos(deg(\x))},{(2.0))*sin(deg(\x))});
\draw[domain=0:2*pi,samples=200,smooth, variable=\x,fill=black!5] plot ({7.2+(1.8+0.01*(1*cos(deg(5*\x))+2*cos(deg(9*\x))+3*cos(deg(18*\x))+4*cos(deg(13*\x))))*cos(deg(\x))},{(1.8+0.01*(1*cos(deg(5*\x))+2*cos(deg(9*\x))+3*cos(deg(18*\x))+4*cos(deg(13*\x))))*sin(deg(\x))});

\draw[domain=0:2*pi,samples=200,smooth, variable=\x] plot ({(2.0)*cos(deg(\x))},{6
+(2.0)*sin(deg(\x))});
\draw[domain=0:2*pi,samples=200,smooth, variable=\x,fill=black!5] plot ({(1.8)*cos(deg(\x))},{6+(1.8)*sin(deg(\x))});

\draw[domain=0:2*pi,samples=200,smooth, variable=\x] plot ({7.2+(2.0)*cos(deg(\x))},{6+((2.0))*sin(deg(\x))});
\draw[domain=0:2*pi,samples=200,smooth, variable=\x,fill=black!5] plot ({7.2+(1.8+0.01*(1*cos(deg(7*\x))+2*cos(deg(10*\x))+3*cos(deg(17*\x))+4*cos(deg(21*\x))))*cos(deg(\x))},{6+(1.8+0.01*(1*cos(deg(7*\x))+2*cos(deg(10*\x))+3*cos(deg(17*\x))+4*cos(deg(21*\x))))*sin(deg(\x))});

\node[] (n1) at (2.1,6) {};
\node[] (n2) at (5.0,6) {};

\node[] (n3) at (0.0,2.1) {};
\node[] (n4) at (0.0,3.8) {};

\node[] (n5) at (7.2,2.1) {};
\node[] (n6) at (7.2,3.8) {};

\node[] (n7) at (2.1,0) {};
\node[] (n8) at (5.0,0) {};

\draw[-latex] (n3) to node[midway, left]{(a)} (n4);
\draw[latex-latex] (n1) to node[midway, sloped, above]{(b)} (n2);
\draw[latex-] (n5) to node[midway, right]{(c)} (n6);
\draw[-latex, thick] (n7) to node[midway, below]{map} (n8);


\node[] at (7.2,0) {$g_{\mu\nu}^{\text{\tiny AdS}}$};
\node[] at (7.2,6) {$g_{\mu\nu}, \theta$};
\node[] at (0,0) {$g_{\mu\nu}^{\text{\tiny AdS}}$};
\node[] at (0,6) {$g_{\mu\nu}^{\text{\tiny AdS}}, \theta$};

\node[] at (0,-2.4) {Base AdS$_2$};
\node[] at (7.2,-2.4) {Target AdS$_2$};


\end{tikzpicture}
	\caption{(a) From the base AdS$_2$ (the exact AdS$_2$ metric, no edge mode), one can one can generate the edge mode on the boundary. (b) Using the (extended) gauge symmetry, one can have generic asymptotic AdS$_2$ metric. (c) The metric can be fixed to be the exact AdS$_2$ metric by the coordinate transformation eating the edge mode.}
	\label{fig: wiggling map}
\end{figure} 

Now we explain the origin of the map from the base AdS$_2$ to the target AdS$_2$ with wiggling boundary in Section~\ref{sec: wiggling boundary}. Let us start with the base AdS$_2$ in the bottom left of Fig.~\ref{fig: wiggling map}. The base AdS$_2$ has the exact AdS$_2$ metric~\eqref{eq: base ads metric} without the edge mode $\theta$. This corresponds to one particular gauge configuration $A_\mu$ in Fig.~\ref{fig: u1 cs description 1}. From the base AdS$_2$, we introduce the edge mode $\theta$, for example, via the gauge transformation by $\theta$ (See (a) in Fig.~\ref{fig: wiggling map}). Hence, the figure on the top left of Fig.~\ref{fig: wiggling map} is analogous to the description by $A_\mu$ and $\theta$ with $U(1)'$ gauge symmetry in Fig.~\ref{fig: u1 cs description 2}. After introducing the edge mode, one can extend the gauge symmetry as well as the metric configuration, which corresponds to the description by $A_\mu,A'_\mu$ and $\phi$ with full $U(1)$ gauge symmetry in Fig.~\ref{fig: u1 cs description 3}. And this gauge symmetry leads to the generic asymptotic AdS$_2$ metric (See (b) in Fig.~\ref{fig: wiggling map}). Finally, one can consider a radial and transverse diffeomorphism similar to Eqs~\eqref{eq: diffeomorphism 1} and \eqref{eq: diffeomorphism 2} to reach the target $AdS_2$ with the exact AdS$_2$ metric (See (c) in Fig.~\ref{fig: wiggling map}). In this transformation, the edge mode $\theta$ is ``eaten'' by the wiggling boundary. And the composite of those procedures corresponds to the map from the base AdS$_2$ to the target AdS$_2$ in Section~\ref{sec: wiggling boundary}.

\section{$sl(2,\mathbb{R})$ BF Theory for Asymptotic AdS$_2$}
\label{sec: sl2 bf gravity}

The framelike formulation of the Jackiw-Teitelboim gravity for AdS$_2$ can be described by $sl(2,\mathbb{R})$ BF theory~\cite{Fukuyama:1985gg,Chamseddine:1989yz,Chamseddine:1989wn}. The action for $sl(2,\mathbb{R})$ BF theory for the two-dimensional manifold $\cM$ is given by
\begin{equation}
	S_{BF}[\Phi,A]\,=\, \int_\cM  \; \tr (\Phi\, F)\ ,
\end{equation}
where $\Phi$ is 0-form $sl(2,\mathbb{R})$ field, and the 2-form field strength $F$ is defined by
\begin{align}
	F\,\equiv\, dA + A\wedge A\ .
\end{align}
Note that unlike Chern-Simons action, $S_{BF}$ without any additional boundary term is gauge invariant. The variation of the action reads
\begin{align}
	\delta S_{BF}\,=\,\mbox{(Bulk E.o.M.)} +\int_{\partial \cM} d\theta \; \tr(\Phi \delta A_\theta)\ .
\end{align}
Hence, if we impose the boundary condition $\delta A=0$, the variational principle is also well-defined. In this case, there is no dynamical degree of freedom left, and the theory is trivial. 

To introduce an edge mode, we add boundary term
\begin{align}
	S_{bdy}\,=\,{\gamma\over 2} \int_{\partial \cM} d\theta \;\tr A_\theta^2\ ,\label{eq: boundary term bf}
\end{align}
which breaks the gauge symmetry on the boundary. Here, $(r,\theta)$ denotes the coordinates for the two-dimensional manifold, and the asymptotic boundary $\partial\cM$ is the surface of constant $r =\infty$. The variation of the total action now becomes
\begin{align}
	\delta S_{tot}\,=\,\delta(S_{BF}+S_{bdy})\,=\, \int_{\partial \cM}d\theta\; \tr [ (\Phi +\gamma A_\theta) \delta A_\theta ]\ ,
\end{align}
and we impose the following boundary condition.
\begin{align}
	\delta A_\theta\big|_{\partial \cM } \,=\, 0\ .
\end{align}
Now, the total action $S_{tot}=S_{BF}+S_{bdy}$ is not invariant under the gauge transformation
\begin{align}
	S_{tot}\quad\longrightarrow\quad  S_{tot} - {\gamma \over 2}  \int_{\partial \cM}d\theta\;  \tr  \bigg[(A_\theta+\partial_\theta h\;h^{-1})^2-A_\theta^2\bigg] \ .
\end{align}
Therefore, for the gauge invariance of the system, we restrict the gauge parameter $h$ on the boundary by
\begin{align}
	h^{-1} \;\partial_\theta h\big|_{\partial \cM } \,=\, 0\ .\label{eq: constant boundary gauge transf}
\end{align}
At the cost of the gauge parameter on the boundary, we have more physical degrees of freedom which would have been gauged by the gauge transformation violating Eq.~\eqref{eq: constant boundary gauge transf}. For example, let us consider two gauge fields $A$ and $\widetilde{A}$ which are related by an ``illegal'' gauge transformation:
\begin{align}
	\widetilde{A}\,=\,h^{-1} A h + h^{-1}d h  \hspace{10mm} \mbox{with}\quad h^{-1} \;\partial_\theta h\big|_{\partial \cM } \ne  0\ .
\end{align}
Without the condition~\eqref{eq: constant boundary gauge transf}, $A$ and $\widetilde{A}$ would have been the identical configuration. But, because we disallowed such a gauge transformation, they are distinct physical configurations.

Note that the boundary term that we added breaks whole boundary gauge symmetry. Hence, we have to restrict boundary gauge symmetry completely. This implies that all configurations with distinct boundary profiles becomes physical, which gives more physical degrees of freedom than the asymptotic AdS solutions that we want to study. Therefore, we revive a part of gauge symmetry by introducing boundary gauge field $\tilde c$.

The $sl(2,\mathbb{R})$ BF theory for the asymptotic AdS$_2$ is defined by
\begin{align}
	S_{AdS}\,=\, S_{BF}(\Phi,\cA) + \int_{\partial \cM} d\theta\;\tr  \left[ {\gamma \over 2} A_\theta^2 - \gamma A_\theta \tilde{c}^{-1} \partial_\theta \tilde{c} \right]\ ,
\end{align}
where $\tilde{c}$ belongs to the nilpotent subgroup 
\begin{align}
	\tilde{c}\,=\, \exp \left[ \begin{pmatrix}
	0 & \lambda\\
	0 & 0\\
	\end{pmatrix}\right]\quad\longrightarrow \quad \tilde{c}^{-1} \partial_\theta \tilde{c}\,=\,\begin{pmatrix}
	0 & \partial_\theta \lambda \\
	0 & 0 \\
	\end{pmatrix}\ .
\end{align}
The gauge field $c$ retrieve a boundary $U(1)$ gauge symmetry. Namely, the action $S_{AdS}$ is invariant under the gauge transformation
\begin{align}
	A\quad &\longrightarrow \quad h^{-1} A h + h^{-1} d h\ ,\label{eq: bf bdy gauge sym 1}\\
	\Phi\quad & \longrightarrow \quad h^{-1} \Phi h\ ,\label{eq: bf bdy gauge sym 2}\\
	\tilde{c}\quad & \longrightarrow \quad \tilde{c}h\ ,\label{eq: bf bdy gauge sym 3}
\end{align}
where $h$ belongs to the nilpotent subgroup:
\begin{align}
	h\,=\, \begin{pmatrix}
	1 & \alpha \\
	0 & 1 \\
	\end{pmatrix}\ .
\end{align}
The recovered $U(1)$ boundary gauge symmetry reduces the boundary physical degree of freedom. As in the Chern-Simons theory, we consider the full boundary gauge transformation $h$, and we promote $h$ to be a physical degrees of freedom $\tilde{g}$. Then, the action for the $SL(2,\mathbb{R})$ BF theory for asymptotic AdS coupled to the gravitational edge mode is 
\begin{align}
	S_{AdS}^{\text{\tiny edge}}\,=\,& S_{BF}(\Phi,\cA) +\int_{\partial \cM} d\theta\; \tr \bigg[\,  {\gamma \over 2}  (A_\theta+ \partial_\theta \tilde{g}\, \tilde{g}^{-1})^2  -\gamma (  \tilde{g}^{-1}A_\theta \tilde{g}  + \tilde{g}^{-1} \partial_\theta \tilde{g} ) \tilde{c}^{-1} \partial_\theta \tilde{c} \, \bigg]\ .\label{eq: bf action with edge mode}
\end{align}
This can be understood as a decomposition of $A$ into the boundary edge degree of freedom and the ``bulk degrees of freedom''. As before, the full boundary gauge symmetry revives at the cost of the introducing the edge mode $\tilde{g}$:
\begin{align}
	A\quad&\longrightarrow \quad \Lambda^{-1}\, A\, \Lambda + \Lambda^{-1} \,d\Lambda\ ,\label{eq: bf full gauge sym 1}\\
	\Phi\quad&\longrightarrow \quad \Lambda^{-1} \,\Phi\, \Lambda\ ,\label{eq: bf full gauge sym 2}\\
	\tilde{g}\quad&\longrightarrow \quad \Lambda^{-1}\,\tilde{g}\;\big|_{\partial \cM}\ ,\label{eq: bf full gauge sym 3}
\end{align}
where $\Lambda\in SL(2,\mathbb{R})$ is defined in the bulk without any restriction on the boundary. This extension of the full $SL(2,\mathbb{R})$ gauge symmetry is analogous to the $U(1)$ Chern-Simons theory with the description in Fig.~\ref{fig: u1 cs description 3}. And the boundary gauge symmetry~\eqref{eq: bf bdy gauge sym 1}$\sim$\eqref{eq: bf bdy gauge sym 3} now becomes the gauge symmetry of $\tilde{g}$ and $\tilde{c}$
\begin{align}
	\tilde{g}\quad & \longrightarrow \quad \tilde{g}\,h \label{eq: gauge sym 2-1}\ ,\\
	\tilde{c}\quad & \longrightarrow \quad \tilde{c}\,h\label{eq: gauge sym 2-2}\ ,
\end{align}
where $h$ belongs to nilpotent subgroup. 

Note that though $\tilde{g}$ and $\tilde{c}$ were introduced in a similar manner, $\tilde{g}$ becomes physical degree of freedom on the boundary while $\tilde{c}$ is to be fixed. Since $\tilde{c}$ belongs to nilpotent subgroup, $\partial_\theta\lambda$ in $\tilde{c}^{-1}\partial_\theta \tilde{c}$ appears in the action linearly. Hence, $\lambda$ plays a role of Lagrangian multiplier.

The variation of the action with the edge mode~\eqref{eq: bf action with edge mode} is found to be
\begin{align}
	\delta S_{AdS}^{\text{\tiny edge}} \,=\,& \int_{\partial \cM}d\theta\; \tr   \big[(\Phi+ \gamma A_\theta + \gamma\partial_\theta\tilde{g}\,\tilde{g}^{-1} - \gamma\tilde{g} \,\tilde{c}^{-1} \partial_\theta\tilde{c} \,\tilde{g}^{-1} ) \delta A_\theta \big] \cr
    & + \gamma\int_{\partial \cM} d\theta\; \tr \big[  \big(  A_\theta + \partial_\theta \tilde{g}\, \tilde{g}^{-1} +[ A_\theta , \tilde{g}\, \tilde{c}^{-1} \partial_\theta \tilde{c}\, \tilde{g}^{-1}] + \partial_\theta ( \tilde{g} \, \tilde{c}^{-1}\partial_\theta \tilde{c} \, \tilde{g}^{-1} )  \big) \delta\tilde{g}\, \tilde{g}^{-1} \big]\cr
    &-\gamma\int_{\partial \cM} d\theta\;\tr\big[(  \tilde{g}^{-1}A_\theta \tilde{g}  +  \tilde{g}^{-1} \partial_\theta \tilde{g} ) \delta (\tilde{c}^{-1} \partial_\theta \tilde{c})  \big]\ . \label{eq: bf variation of action with edge mode}
\end{align}
The first line of the variation~\eqref{eq: bf variation of action with edge mode} imposes the boundary condition\footnote{This boundary condition (and the boundary term~\eqref{eq: boundary term bf}) is not unique. In Appendix~\ref{app: bc2}, we analyze another boundary condition, $\delta\big(\Phi-\gamma A_\theta\big)=0$.} of the bulk field $A_\theta$
\begin{align}
    \delta A_\theta\big|_{\partial \cM} \,=\,0\ . \label{eq: boundary condition for a naive}
\end{align}
The second line corresponds to the equation of motion of the edge mode $g$. We have observed that the variation of the action leads to the boundary condition of the bulk field and the equation of motion of the edge mode in the $U(1)$ Chern-Simons theory and the metric-like formulation of JT gravity.\footnote{In the literature, the $sl(2,\mathbb{R})$ BF theory does not separate the edge mode from the bulk field. Hence, it corresponds to the description in Fig.~\ref{fig: u1 cs description 1}. In this case, as we have pointed out in Section~\ref{sec: u1 cs theory}, one should be careful in connecting the variation of the action with the boundary condition.} Furthermore, the third line gives the constraint imposed by the Lagrangian multiplier $\partial_\theta\lambda$.

After fixing gauge\footnote{The $sl(2,\mathbb{R})$ generator $L_n$ ($n=0,\pm 1$) is defined by 
\begin{align}
    L_{1}\,=\, \begin{pmatrix}
        0 & 0 \\
        1 & 0\\
    \end{pmatrix}\;\;,\quad L_{0}\,=\, \begin{pmatrix}
        {1\over 2} & 0 \\
        0 & -{1\over 2}\\
    \end{pmatrix}\;\;,\quad L_{1}\,=\, \begin{pmatrix}
        0 & -1 \\
        0 & 0\\
    \end{pmatrix}
\end{align}
}
\begin{align}
	A_r\,=\, b^{-1}(r) \partial_r b(r)\hspace{8mm} \mbox{where} \quad b(r)= e^{rL_0}\ ,
\end{align}
we can rewrite $A_\theta$ and $\tilde{g}$ as
\begin{align}
    \Phi\,=\,& b^{-1} \,\phi\, b\ ,\\
	A_\theta\,=\,& b^{-1} \,a_\theta\, b\ ,\\
	\tilde{g}\,=\, & b^{-1}\, g\, b\ ,\\
	\tilde{c}\,=\, & c\, b\ .
\end{align}
Because of the boundary condition~\eqref{eq: boundary condition for a naive}, we should fix all component of the matrix $a_\theta$. 
Strictly speaking, the boundary condition~\eqref{eq: boundary condition for a naive} should be understood within the trace, \ie\; $\tr\big[(\cdots)\delta A_\theta ]\big|_{\partial \cM}=0$. For example, for a matrix $\delta N$ given by
\begin{align}
    \delta N\,\equiv\,\begin{pmatrix}
        {1\over2} \delta N_0 & - \delta N_{-1} e^{-r} \\
        \delta N_1 e^{r} & -{1\over 2} \delta N_0
    \end{pmatrix}\ ,
\end{align}
the component $\delta N_{-1}$ vanishes as $r\to\infty$. But in the trace of the product with other matrix, it does not vanish. \ie
\begin{align}
    \tr \bigg[\begin{pmatrix}
        {1\over2} M_0 & - M_{-1} e^{-r} \\
        M_1 e^{r} & -{1\over 2} M_0
    \end{pmatrix}\begin{pmatrix}
        {1\over2} \delta N_0 & - \delta N_{-1} e^{-r} \\
        \delta N_1 e^{r} & -{1\over 2} \delta N_0
    \end{pmatrix} \bigg]\,=\, {1\over 2}M_0 \delta N_0 - M_{-1}\delta N_1 - M_1\delta N_{-1}\ .
\end{align}
This is the main reason why we fix all the component of the matrix $A_\theta$ and why we introduced the edge mode $g$ separately on top of the $A$ and $\Phi$.

Using the full gauge symmetry~\eqref{eq: bf full gauge sym 1}$\sim$\eqref{eq: bf full gauge sym 3}, one can fix $a_\theta$ to be the following constant $sl(2,\mathbb{R})$ matrix.
\begin{align}
    a_\theta\,=\, \kappa\begin{pmatrix}
    0 & -\cL_0\\
    1 & 0\\
    \end{pmatrix}\ ,\label{eq: atheta sol}
\end{align}
where $\cL_0$ is a constant. Note that $\Phi$ is the Lagrangian multiplier of the bulk. Therefore, it can be fixed by the equation of motion for $\Phi$:
\begin{align}
    d\Phi + [A,\Phi]\,=\, 0\ ,
\end{align}
The solution for $\Phi$ can be found to be 
\begin{align}
    \Phi \,=\, b^{-1}\phi(\theta) b\,=\, b^{-1}\,e^{-a_\theta \theta} \,\phi_0 \, e^{a_\theta \theta}\,b\ ,
\end{align}
where $\phi_0$ is a constant $sl(2,\mathbb{R})$ matrix. \ie
\begin{align}
    \phi_0 \,=\, \begin{pmatrix}
    {1\over 2}\chi_0 & -\chi_{-1} \\
    \chi_1 & -{1\over 2}\chi_0  \\
    \end{pmatrix}\ .
\end{align}
The constraint imposed by $c^{-1}\partial_\theta c$ reads 
\begin{align}
	(g^{-1} a_\theta g + g^{-1}\partial_\theta g )_1\quad:\quad \mbox{constant}\ . \label{eq: bf constraint}
\end{align}
Here, $(M)_a$ ($a=0,\pm 1$) denotes each component of a $sl(2,\mathbb{R})$ element $M$:
\begin{align}
M=\begin{pmatrix} {1\over 2}(M)_0 & -(M)_{-1}\\
(M)_1 & -{1\over 2}(M)_0\\
\end{pmatrix}\ .
\end{align}
The constant~\eqref{eq: bf constraint} is fixed to be $\kappa$ because $g$ includes the identity matrix. 
\begin{align}
	(g^{-1} a_\theta g + g^{-1}\partial_\theta g )_1\,=\,\kappa\ . \label{eq: constraint}
\end{align}
Using the residual gauge symmetry given in \eqref{eq: gauge sym 2-1} and \eqref{eq: gauge sym 2-2}, we choose the gauge condition
\begin{align}
	(g^{-1} a_\theta g + g^{-1}\partial_\theta g )_0\,=\,0\ .\label{eq: gauge condition}
\end{align}
To see the remaining physical degrees of freedom for $g$, we parametrize $SL(2,\mathbb{R})$ matrix $g$ by
\begin{align}
    g^{-1} a_\theta g + g^{-1}\partial_\theta g \,=\, k^{-1}\partial_\theta k\ . \label{eq: parametrization a}
\end{align}
Note that $g^{-1} a_\theta g + g^{-1}\partial_\theta g = k^{-1}\partial_\theta k$ does not distinguish $k$ and $-k$. Hence, $k$ belongs to $PSL(2,\mathbb{R})$ by identifying $k$ with $-k$. To solve the constraint and the gauge condition, we will use the Iwasawa decomposition of $k$ given by
\begin{align}
    k(\theta)\,=\,\begin{pmatrix}
	\cos { \Omega\, u(\theta)\over 2} & -\sin {\Omega \,u(\theta)\over 2}\\
	\sin {\Omega \,u(\theta)\over 2} & \cos {\Omega \, u(\theta)\over 2}\\
	\end{pmatrix}\begin{pmatrix}
	[y(\theta)]^{-{1\over 2}} & 0\\
	0 & [y(\theta)]^{1\over 2} \\
	\end{pmatrix}\begin{pmatrix}
	1 & f(\theta) \\
	0 & 1\\
	\end{pmatrix}\ .\label{eq: iwasawa decomposition}
\end{align}
where $y(\theta)$ and $f(\theta)$ is periodic with period $\beta$, and $u(\theta)$ has winding number 1. \ie
\begin{align}
    y(\theta+\beta)\,=\, y(\theta)\;\;,\quad f(\theta+\beta)\,=\, f(\theta) \;\;,\quad u(\theta+\beta)\,=\, u(\theta) +\beta\ .
\end{align}
In this work, we demand that $g^{-1} a_\theta g + g^{-1}\partial_\theta g = k^{-1}\partial_\theta k$ is single-valued. \ie
\begin{align}
    k(\beta)\,=\, \pm k(0)\ .
\end{align}
This gives us the relation between $\beta$ and $\Omega$:
\begin{align}
	{\Omega\over 2} \,=\, {n \pi\over \beta}  \hspace{10mm} \mbox{where}\quad n\in \mathbb{Z}\ .\label{eq: single-value condition}
\end{align}
And the periodic condition for $k(\theta)$ reads
\begin{align}
    k(\theta+\beta)\,=\, (-1)^n k(\theta)\ .
\end{align}
The global structure can also be seen in the following holonomy along the Euclidean time $\theta$.
\begin{align}
    \hol \big(\tilde{g}A\tilde{g}^{-1}-\tilde{g}d\tilde{g}\big)\,\equiv&\, \cP \exp \bigg[\oint_{\partial \cM} \big(\tilde{g}A\tilde{g}^{-1}-\tilde{g}d\tilde{g}\big)\bigg]\,\sim \, \exp\big(\beta a_\theta\big)\cr
    \sim & \, k(\beta)\, k^{-1}(0)\,=\, (-1)^n\begin{pmatrix}
    1 & 0 \\
    0 & 1 \\
    \end{pmatrix}\ .
\end{align}
The single-value condition~\eqref{eq: single-value condition} makes the holonomy trivial. Namely, it belongs to the center subgroup $\{\pm I\}$ of $SL(2,\mathbb{R})$. Furthermore, it determines $\cL_0$ in $a_\theta$ in terms of $\beta$. 
\begin{align}
    \kappa \sqrt{\cL_0}\,=\, {n\pi \over \beta}\ .
\end{align}
This trivial holonomy condition can be interpreted as the smoothness of the geometry~\cite{Castro:2011iw}. From now on, let us focus on $n=1$ case where $k(\theta)$ is a map from $S^1$ to $PSL(2,\mathbb{R})$ with winding number 1.

Using the Iwasawa decomposition~\eqref{eq: iwasawa decomposition}, one can determine $y(\theta)$ and $f(\theta)$ from the constraint~\eqref{eq: constraint} and the gauge condition~\eqref{eq: gauge condition}, respectively.
\begin{align}
    y(\theta)\,=\,& {\Omega u' \over 2 \kappa}\ ,\\
    f(\theta)\,=\,& -{u''\over 2\kappa u'} \ .
\end{align}
and we have
\begin{align}
   g^{-1} a_\theta g + g^{-1}\partial_\theta g \,=\,\begin{pmatrix}
        0 & -{1\over2\kappa}\bigg[Sch[u,\theta] + {1\over 2}\Omega^2u'^2\bigg]\\
        \kappa & 0\\
    \end{pmatrix}\ .
\end{align}
Therefore the BF action with edge mode~\eqref{eq: bf action with edge mode} becomes
\begin{align}
    S_{AdS}^{\text{\tiny edge}}\,=\, \gamma \int d\theta\, \tr\big(g^{-1} a_\theta g + g^{-1}\partial_\theta g \big)^2\,=\, -\gamma \int d\theta \, \bigg[Sch[u,\theta] + {2\pi^2\over \beta^2}u'^2\bigg]\ .\label{eq: bf schwarzian action}
\end{align}
where we used the single-value condition~\eqref{eq: single-value condition} with $n=1$. 

The measure of the edge mode can be derived from the Haar measure of $PSL(2,\mathbb{R})$~\cite{Cotler:2018zff}. 
We use the Iwasawa decomposition~\eqref{eq: iwasawa decomposition}, where the Haar measure is given by
\begin{align}
    \cD \mu \,=\, {\cD u \,\cD y \,\cD f\over {\displaystyle \prod_\theta}[y(\theta)]^2} \ .
\end{align}
Together with the constraint and the gauge condition, we find
\begin{align}
    &\int {\cD u \,\cD y \,\cD f\over {\displaystyle \prod_\theta}[y(\theta)]^2}\, \delta \big((a_\theta)_1-\kappa \big)\delta \big((a_\theta)_0\big)\,=\,\int {\cD u \,\cD y \,\cD f\over {\displaystyle \prod_\theta} [y(\theta)]^2}\, \delta \bigg({\Omega u'\over y}-\kappa \bigg)\delta \big((a_\theta)_0\big)\\
    \,=\,&\int {\cD u \,\cD f\over{\displaystyle \prod_\theta}u'(\theta)}\,\delta \bigg(-\kappa f-{u''\over 2u'}\bigg)\,=\,  \int {\cD u \over {\displaystyle \prod_\theta}u'(\theta)}\,,
\end{align}
and hence recover the result of \cite{Cotler:2018zff}
which has been obtained in the Gauss decomposition.

The parameterization of $g^{-1}a_\theta g g^{-1}\partial_\theta g$ by $k$ in Eq.~\eqref{eq: parametrization a} is invariant under $k\to \Upsilon_0 k$ where $\Upsilon _0$ is a constant $PSL(2,\mathbb{R})$ element. 
\begin{align}
    \Upsilon_0\,\equiv \, \begin{pmatrix}
        d & c\\
        b & a\\
    \end{pmatrix}\hspace{10mm}\mbox{where}\quad ad-bc\,=\,1\ .
\end{align}
Therefore, this redundant description should be eliminated by imposing the equivalent relation
\begin{align}
    \Upsilon_0\, k\,\sim \, k \ .\label{eq:equivalence rel of k}
\end{align}
This equivalence relation can be translated into the equivalence relation of $u(\theta)$ via the Iwasawa decomposition~\eqref{eq: iwasawa decomposition}:
\begin{align}
    {a \tan {\pi u(\theta) \over \beta} + b \over c \tan {\pi u(\theta) \over \beta}+d}\,\sim \, \tan {\pi u(\theta) \over \beta} \ .
\end{align}
This equivalence relation becomes the $PSL(2,\mathbb{R})$ gauging of the edge mode action~\eqref{eq: bf schwarzian action}. We have seen in Section~\ref{sec: jt gravity for nearly ads2} that the $PSL(2,\mathbb{R})$ gauging comes from the isometry of the (exact) AdS$_2$. In the BF formulation, we can also confirm that it corresponds to the isometry of the (background) bulk field $a_\theta$. To see this, we first obtain the relation\footnote{See Appendix~\ref{app: iwasawa and gauss} for the explicit relation.} between $k$ and $g$ from Eq.~\eqref{eq: parametrization a}.
\begin{align}
    k\,=\, e^{a_\theta \theta}\, g \ ,
\end{align}
where $a_\theta$ is fixed to be a constant element in Eq.~\eqref{eq: atheta sol}. Then, the equivalence relation~\eqref{eq:equivalence rel of k} leads to the equivalence relation of $g$:
\begin{align}
    \Upsilon \, g\,\sim \, g\ , \label{eq: equiv relation of g}
\end{align}
where $\Upsilon (\theta)$ is defined by
\begin{align}
	 \Upsilon\,\equiv\, e^{-a_\theta \theta }\,\Upsilon_0\, e^{a_{\theta }  \theta}\ . \label{eq: upsilon def}
\end{align}
One can easily see that this is a general solution of the following equation.
\begin{align}
    \Upsilon^{-1} a_\theta \Upsilon + \Upsilon^{-1}\partial_\theta \Upsilon \,=\,a_\theta\ .\label{eq: bf isometry}
\end{align}
This equation means the redundant parametrization of $g$ by $\Upsilon$, and therefore, we can again confirm that the equivalence relation of $g$~\eqref{eq: equiv relation of g} should be imposed. Moreover, in Eq.~\eqref{eq: bf isometry}, $\Upsilon$ can be interpreted as a gauge transformation parameter which does not change the background bulk gauge field $a_\theta$, in other words, the isometry of the background AdS$_2$.

\section{Conclusion}
\label{sec: conclusion}

In this work, we have studied the gravitational edge mode in the JT gravity and the $sl(2,\mathbb{R})$ BF theory with the asymptotic AdS$_2$ boundary condition. We revisited the derivation of the Schwarzian action from the wiggling boundary of the JT gravity. We introduced a new counter term $K_0$ which plays an important role in obtaining the correct Schwarzian action. We discussed the variation of the action which is involved not only with the boundary condition of the bulk metric and the bulk dilaton but also with the equation of motion of the gravitational edge mode. Introducing the target and the base AdS$_2$, we showed that the inversion between the base and the target AdS$_2$ gives the Schwarzian action. In addition, we demonstrated that this inversion naturally leads to the path integral measure for the Schwarzian theory. We explicitly showed that the redundant description of the base AdS$_2$ corresponding to the isometry induces the $PSL(2,\mathbb{R})$ gauging of the Schwarzian action. With the boundary of constant AdS radial coordinate without wiggling, we showed that the broken radial diffeomorphism leads to the would-be gauge mode, in other words, the gravitational edge mode. We demonstrated that this gravitational edge mode in the asymptotically AdS$_2$ can be described by the Schwarzian action. We also presented the relation between the gravitational edge mode and the wiggling boundary.

In the $sl(2,\mathbb{R})$ BF theory, we incorporated the edge mode in the BF theory based on the $U(1)$ Chern-Simons example. We clarified the variation of the action and the corresponding boundary condition of the bulk field. We demonstrated that the Haar measure of the Iwasawa decomposition of the $SL(2,\mathbb{R})$ gives the path integral measure of the Schwarzian theory. We also showed that the redundancy in the Iwasawa decomposition, which is equivalent to the isometry of the AdS$_2$ background, brings about the $PSL(2,\mathbb{R})$ gauging of the Schwarzian action.

The overall sign of our JT gravity action~\eqref{eq: jt action} is opposite to that of the literature for ``nearly-AdS''. We chose this sign convention to have the positive dilaton solution\footnote{The dilaton $\phi$ plays a role of ``area'' in the Bekenstein-Hawking entropy of the AdS$_2$, and this is the reason for our choice of the positive dilaton solution. However, strictly speaking, since one can flip the sign of the dilaton $\phi\to -\phi$, we could have begun with the other sign convention of the action if we allow the negative dilaton solution. In this case, the ``area'' is related to $-\phi$.} $\phi$ and the stability of the Schwarzian action. One can easily expect the opposite sign from the inversion formula of the Schwarzian derivative~\eqref{eq: inversion formula of schwarzian derivative}. From the point of view of the higher dimensional near extremal black hole~\cite{Sarosi:2017ykf,Nayak:2018qej,Nayak:2018qej}, the dilaton of the JT gravity originates from the transverse area of the fixed radial hypersurface. After the dimensional reduction to AdS$_2$, the JT action with our convention~\eqref{eq: jt action} can be obtained by expanding the dilaton $\Phi$ coming from the higher dimension around the constant $\phi_0$, which is related to the entropy of the higher dimensional extremal black hole, as follows.
\begin{align}
    \Phi^2\,=\, \phi_0 - \phi\ .
\end{align}
As a result, the physical requirement for the stability of the near extremal black hole translates to
$\Phi^2$ becoming smaller as the black hole deviates from extremality. 

It would be interesting to see if there are any interesting consequences of the inversion of degree of freedom that we encountered, namely from $\theta(u)$ to $u(\theta)$. Perhaps the natural place to look for is in how the Schwarzian modes couples to other matter fields. In this context, one might have to revisit the calculation of correlation functions. Nevertheless, we expect that some perturbative calculations would still hold, because the perturbative expansion around the classical solution $u_{cl}(\theta)=\theta$ of the Schwarzian action gives the same perturbative expansion of the inverse function up to order $\cO(\epsilon)$ (and up to sign). \ie
\begin{align}
    u(\theta)\,=\, \theta +\epsilon(\theta) \;\;\longrightarrow \;\; \theta(u)\,=\, u - \epsilon (u) +\cO(\epsilon^2)\ .
\end{align}
For example, the leading contribution of the Schwarzian mode to the four point function of the matter might not be able to see the difference. We leave this issue for future works.


\vspace{3mm}

\paragraph{Acknowledgements.} 

We would like to thank Ioannis Papadimitriou, Piljin Yi, Akhil Sivakumar and Frank Ferrari for discussions. P.N. and J.Y. would like to thank International Centre for Theoretical Sciences (ICTS), Tata institute of fundamental research, Bengaluru for hospitality and support during the initial stage of this work. 
The work of E.J. was supported by the National Research Foundation of Korea (NRF) grant funded by the Korea government (MSIT) (No. 2022R1F1A1074977).
The work of J.Y. was supported by the NRF grant funded by the Korea government (MSIT) (No. 2022R1A2C1003182). J.Y. is supported by an appointment to the JRG Program at the APCTP through the Science and Technology Promotion Fund and Lottery Fund of the Korean Government. J.Y. is also supported by the Korean Local Governments - Gyeongsangbuk-do Province and Pohang City. P.N would like to acknowledge SERB grant
MTR/2021/00014.

\appendix

\section{Other Boundary Condition for BF Theory}
\label{app: bc2}

In this appendix, we repeat the same calculations as in Section~\ref{sec: sl2 bf gravity} with a different boundary term given by
\begin{align}
	S_{bdy}\,=\,  -\int_{\partial \cM} d\theta\; \tr\big(\Phi A_\theta\big) + {\gamma \over 2} \int_{\partial \cM} d\theta\; \tr \big(A_\theta^2\big)\ .
\end{align}
The variation of the total action $S_{BF}+S_{bdy}$ becomes
\begin{align}
	\delta S_{tot}\,\equiv\,\delta(S_{BF}+S_{bdy})= -  \int_{\partial \cM}d\theta\; \tr [A \delta (\Phi-\gamma A_\theta) ]\ .
\end{align}
Hence, one can choose the boundary condition for the bulk field to be
\begin{align}
	\delta \big(\Phi-\gamma  A_\theta\big)\bigg|_{\partial \cM}\,=\,0\ . 
\end{align}
In the same way as in Section~\ref{sec: sl2 bf gravity}, we 
\begin{align}
	S_{AdS}^{\text{\tiny edge}}=& S_{BF}(\Phi,A) + \int_{\partial \cM} d\theta\; \tr \bigg[- \Phi (A_\theta +  \partial_\theta\tilde{g}\, \tilde{g}^{-1})  + {\gamma \over 2} (A_\theta+ \partial_\theta\tilde{g}\, \tilde{g}^{-1})^2  \bigg] \cr
	&+ \int d\theta\; \tr\bigg[(\tilde{g}^{-1}\,\Phi\, \tilde{g} - \gamma \,\tilde{g}^{-1}A_\theta \tilde{g}  - \gamma \,\tilde{g}^{-1} \partial_\theta \tilde{g})\, \tilde{c}^{-1} d\tilde{c} \bigg]\ ,
\end{align}
where $\tilde{c}$ is an element in the nilpotent subgroup 
\begin{align}
	\tilde{c}\,=\, \exp \left[ \begin{pmatrix}
	0 & \lambda(\theta)\\
	0 & 0\\
	\end{pmatrix}\right]\ .
\end{align}
This action is invariant under the gauge symmetry
\begin{align}
	A\quad&\longrightarrow \quad \Lambda^{-1} \,A\, \Lambda + \Lambda^{-1}\, d\Lambda\ ,\\
	\Phi\quad&\longrightarrow \quad \Lambda^{-1} \,\Phi\, \Lambda\ ,\\
	\tilde{g}\quad&\longrightarrow \quad \Lambda^{-1}\, \tilde{g}\ ,
\end{align}
and
\begin{align}
	\tilde{g}\quad & \longrightarrow \quad \tilde{g}\,h\ ,\\
	\tilde{c}\quad & \longrightarrow \quad \tilde{c}\,h\ ,
\end{align}
where $\Lambda\in SL(2,\mathbb{R})$ and $h$ belongs to the nilpotent subgroup. The gauge transformation by $\Lambda\in SL(2,\mathbb{R})$ comes from the ambiguity of the decomposition, and choosing the reference point for the decomposition corresponds to fixing this gauge symmetry. The residual gauge symmetry will play a role of $PSL(2,\mathbb{R})$ gaugeing of the Schwarzian action.

As before, we fix the gauge 
\begin{align}
	A_r \,=\, b^{-1} \partial_r b \hspace{8mm} \mbox{where} \quad b(r)= e^{rL_0}\ ,
\end{align}
and we define 
\begin{align}
	\Phi\,=\,& b^{-1}\, \phi \,b\ ,\\
	A_\theta\,=\,& b^{-1}\, a_\theta\,b\ ,\\
	\tilde{g}\,=\,&b^{-1}\, g \,b\ ,\\
	\tilde{c}\,=\,&b^{-1}\, c\, b\ .
\end{align}
In addition, by using the gauge symmetry, one can further fix $\bar{a}_\theta\equiv a_\theta-{1\over \gamma \phi}$ to be constant.
\begin{align}
	\bar{a}_\theta\,\equiv\, a_\theta -{1\over \gamma }\phi \,=\, \kappa\,\begin{pmatrix}
	0 & -\cL_0\\
	1 & 0 \\
	\end{pmatrix}\ .
\end{align}
where $\cL_0$ and $\kappa$ is a constant. This is consistent with the boundary condition.
\begin{align}
	 A-{1\over \gamma}\Phi\quad:\quad \mbox{fixed}\ .
\end{align}
The bulk dilaton, as a Lagrangian multiplier, can be determinant by the equation of motion
\begin{align}
    \partial_\theta \phi(\theta) + [a_\theta,\phi]\,=\, \partial_\theta \phi(\theta) + [\bar{a}_\theta,\phi]\,=\,0 \ .
\end{align}
The solution for the dilaton $\phi $ can be written as
\begin{align}
    \phi \,=\, e^{-\bar{a}_\theta \theta}\, \phi_0 \,e^{\bar{a}_\theta \theta}\ ,
\end{align}
where $\phi_0$ is a constant $sl(2,\mathbb{R})$ matrix. 
\begin{align}
    \phi_0 \,=\, \begin{pmatrix}
    {1\over 2}\chi_0 & -\chi_{-1} \\
    \chi_1 & -{1\over 2}\chi_0  \\
    \end{pmatrix}\ ,
\end{align}
and therefore, $a_\theta$ is given by
\begin{align}
    a_\theta\,=\, \begin{pmatrix}
    {1\over 2\gamma}\chi_0 & -\kappa\cL_0-{1\over \gamma}\chi_{-1} \\
    \kappa+{1\over \gamma}\chi_1  & -{1\over 2\gamma}\chi_0  \\        
    \end{pmatrix}\ .
\end{align}

The constraint imposed by the Lagrangian multiplier $\partial_\theta \lambda$ is 
\begin{align}
	\big( g^{-1}\,\bar{a}_\theta\, g +  g^{-1}\, \partial_\theta g \big)_1\,=\, \kappa\ .
\end{align}
And we choose the gauge condition from the residual gauge symmetry
\begin{align}
	\big( g^{-1}\,\bar{a}_\theta\, g +  g^{-1}\, \partial_\theta g \big)_0\,=\,0\ .
\end{align}
To solve the constraint and the gauge condition, let us consider the parametrization of $g^{-1}\, \bar{a}_\theta \, g + g^{-1}\partial_\theta g$ in terms of $k\in PSL(2,\mathbb{R})$.
\begin{align}
    g^{-1}\, \bar{a}_\theta\, g + g^{-1}\partial_\theta g \,=\, k^{-1}\,\partial_\theta k\ . \label{eq: parametrization abar}
\end{align}
We consider the Iwasawa decomposition of $k$ given by
\begin{align}
    k(\theta)\,=\,\begin{pmatrix}
	\cos { \Omega\, u(\theta)\over 2} & -\sin {\Omega \,u(\theta)\over 2}\\
	\sin {\Omega \,u(\theta)\over 2} & \cos {\Omega \, u(\theta)\over 2}\\
	\end{pmatrix}\begin{pmatrix}
	[y(\theta)]^{-{1\over 2}} & \\
	0 & [y(\theta)]^{1\over 2} \\
	\end{pmatrix}\begin{pmatrix}
	1 & f(\theta) \\
	0 & 1\\
	\end{pmatrix}\ .
\end{align}
where $y(\theta)$ and $f(\theta)$ is a periodic function of $\theta$ with period $\beta$ while $u(\theta)$ is a function of $\theta$ with winding number 1. \ie
\begin{align}
    y(\theta+\beta)\,=\, y(\theta)\;\;,\quad f(\theta+\beta)\,=\, f(\theta) \;\;,\quad u(\theta+\beta)\,=\, u(\theta) +\beta\ .
\end{align}
In the same way as in Section~\ref{sec: sl2 bf gravity}, we have
\begin{align}
	S_{AdS}^{\text{\tiny edge}}\, =\,- {\gamma \over 2}\int d\theta\, \left[  Sch[u,\theta] +{1 \over 2} u^2 \theta'^2 \right] - {\beta\over \gamma }\bigg({1\over 4}\chi_0^2 -\chi_1\chi_{-1}\bigg)\ .
\end{align}
One can determine $\Omega$ from the global structure. In this work, we demand that $g^{-1}\bar{a}_\theta\, g +  g^{-1} \partial_\theta g$ is a single-valued function, in other words, the holonomy of $g^{-1}\bar{a}_\theta\, g +  g^{-1} \partial_\theta g$ along the Euclidean time is trivial. Then, we find
\begin{align}
    \kappa \sqrt{\cL_0}\,=\, {n \pi \over \beta} \,=\,{\Omega\over 2}\ .
\end{align}
And for $n=1$, the action becomes
\begin{align}
	S_{AdS}^{\text{\tiny edge}}\, =\,- {\gamma \over 2}\int d\theta\, \left[  Sch[u,\theta] +{2\pi^2 \over \beta^2}u'^2 \right] - {\beta\over \gamma }\bigg({1\over 4}\chi_0^2 -\chi_1\chi_{-1}\bigg)\ .
\end{align}
In this case, although the holonomy $\hol(\bar{a}_\theta)$ is trivial (\ie\; it belongs to the center subgroup), $\hol(a_\theta)$ is not.

\section{Iwasawa and Gauss Decomposition}
\label{app: iwasawa and gauss}

In this appendix, we present the explicit relation between the Iwasawa decomposition of $k(\theta)$ and the Gauss decomposition of $g(\theta)$ in the following equation.
\begin{align}
    g^{-1}\, a_\theta\, g + g^{-1}\partial_\theta g \,=\, k^{-1}\,\partial_\theta k\ ,
\end{align}
where the constant matrix $a_\theta$ is given by
\begin{align}
    a_\theta \,=\, \kappa\,\begin{pmatrix}
        0 & -\cL_0\\
        1 & 0\\
    \end{pmatrix}\ .
\end{align}
The Iwasawa decomposition is given by
\begin{align}
    k(\theta)\,=\,\begin{pmatrix}
	\cos { \Omega\, u(\theta)\over 2} & -\sin {\Omega \,u(\theta)\over 2}\\
	\sin {\Omega \,u(\theta)\over 2} & \cos {\Omega \, u(\theta)\over 2}\\
	\end{pmatrix}\begin{pmatrix}
	[y(\theta)]^{-{1\over 2}} & \\
	0 & [y(\theta)]^{1\over 2} \\
	\end{pmatrix}\begin{pmatrix}
	1 & f(\theta) \\
	0 & 1\\
	\end{pmatrix}\ .
\end{align}
where $y(\theta)$ and $f(\theta)$ is a periodic function with period $\beta$, and $u(\theta)$ is a function of $\theta$ with winding number 1. The Gauss decomposition of $g(\theta)$ is 
\begin{align}
	g\,=\,\begin{pmatrix}
	1 & 0 \\
	E(\theta) & 1\\
	\end{pmatrix}\begin{pmatrix}
	[\Lambda(\theta)]^{-{1\over 2}} & \\
	0 & [\Lambda(\theta)]^{1\over 2}  \\
	\end{pmatrix}\begin{pmatrix}
	1 & F(\theta)\\
	0 & 1\\
	\end{pmatrix} \ ,
\end{align}
where $E(\theta), \Lambda(\theta)$ and $F(\theta)$ is a periodic function with period $\beta$. We find
\begin{align}
     u \,=\, &{2\over \Omega}\arctan\bigg[{ E(\theta) \cos \big(\kappa\sqrt{\cL_0}\theta\big) + {1\over \sqrt{\cL_0} }\sin \big(\kappa\sqrt{\cL_0}\theta\big) \over \cos \big(\kappa\sqrt{\cL_0}\theta\big)-E(\theta) \sqrt{\cL_0} \sin \big(\kappa\sqrt{\cL_0}\theta\big)}\bigg]\ ,\\
    {1\over y}\,=\, &  {1-\cL_0 \over 2\cL_0 \Lambda(\theta)}\bigg[   (\cL_0 [E(\theta)]^2-1)\cos\big(2\kappa\sqrt{\cL_0}\theta\big)+2\cL_0 E(\theta) \sin \big(2\kappa\sqrt{\cL_0}\theta\big)  \bigg]\cr
    &+{(1+\cL_0)(1+\cL_0 [E(\theta)]^2)\over 2\cL_0 \Lambda(\theta)}\ ,\\
    {f\over y}\,=\,& {(1-\cL_0)\over 2\cL_0 \Lambda}\bigg[\big( (\cL_0E\Lambda +F(\cL_0 E^2 -1))\cos\big(2\kappa\sqrt{\cL_0}\theta\big) + \sqrt{\cL_0} (2EF+\Lambda)\sin\big(2\kappa\sqrt{\cL_0}\theta\big) \big)\bigg]\cr
    &+{1\over 2\cL_0 \Lambda}(1+\cL_0)(F+\cL_0E^2F+\cL_0 E\Lambda) \ .
\end{align}
Under this transformation, the Haar measure of the Gauss decomposition can be obtained from that of the Iwasawa decomposition:
\begin{align}
    \int {\cD u \,\cD y \,\cD f\over {\displaystyle \prod_\theta}[y(\theta)]^2}\,=\,\int {\cD E \,\cD \Lambda \,\cD F\over {\displaystyle \prod_\theta}[\Lambda(\theta)]^2} \ .
\end{align}

\bibliographystyle{JHEP}
\bibliography{bfgravity}

\end{document}